\documentstyle[prb,aps]{revtex}
\begin{document}
\input{epsf.tex}
\twocolumn
\title{Specific Heat (1.2--108\,K) and Thermal Expansion (4.4--297\,K)
Measurements of the 3$d$ Heavy Fermion Compound LiV$_2$O$_4$}
\author{D. C. Johnston, C. A. Swenson, and S. Kondo}
\address{Ames Laboratory and Department of Physics and Astronomy, Iowa State
University, Ames, Iowa 50011}
\date{Phys.\ Rev.\ B, to be published.}
\maketitle
\begin{abstract}Specific heat $C_{\rm p}(T)$ measurements of the heavy fermion
normal-spinel structure compound LiV$_2$O$_4$ were carried out using a heat-pulse
calorimeter over the temperature $T$ range from 1.2 to 108 K\@.  The electronic
specific heat $C_{\rm e}(T)$ of LiV$_2$O$_4$ is extracted from the $C_{\rm p}(T)$
data using the lattice contribution obtained for LiTi$_2$O$_4$, a superconductor
with $T_{\rm c}$ = 11.8\,K\@. The electronic specific heat coefficient
$\gamma(T)\equiv C_{\rm e}(T)/T$ of LiV$_2$O$_4$ is  found to be 0.42 and
0.43\,J/mol\,K$^2$ at $T = 1$\,K for two different high magnetic purity samples,
respectively.  $\gamma(T)$ decreases rapidly with increasing temperature from 4
to 30\,K and then decreases much more slowly from 0.13\,J/mol\,K$^2$ at 30\,K to
0.08\,J/mol\,K$^2$ at 108\,K\@.  The $C_{\rm e}(T)$ of the first of the above
two LiV$_2$O$_4$ samples is compared with  theoretical predictions for the spin
$S = 1/2$ Kondo model, a generic Fermi liquid model, and an
antiferromagnetically coupled quantum-disordered metal.  Each of these theories
can adequately describe the $T$ dependence of $C_{\rm e}$ in the Fermi liquid
regime at low ($\sim$\, 1--10\,K) temperatures, consistently yielding a large
extrapolated $\gamma(0)$ = 428(3)\,mJ/mol\,K$^2$.  However, none of these
theories describes $C_{\rm e}(T)$ from $\sim 10$\,K to 108\,K\@.  Our $C_{\rm
e}(T)$ data are also in severe disagreement with the magnetic specific heat of
the spin $S = 1/2$ Heisenberg model, calculated above $\sim 40$\,K for the V
sublattice of the spinel structure.  Thermal expansion measurements of
LiV$_2$O$_4$ were carried out from 4.4 to 297\,K using a differential
capacitance dilatometer.  Strong increases in the thermal expansion coefficient
and Gr\"uneisen parameter $\Gamma$ are found below $\sim 20$\,K, confirming the
results of  Chmaissem {\it et al.}\ [Phys.\ Rev.\ Lett.\ {\bf 79}, 4866 (1997)]
obtained using neutron diffraction.  We estimate  $\Gamma(0)\approx 11.4$, which
is intermediate between those of conventional metals and $f$-electron heavy
fermion compounds. 
\end{abstract}

\section{Introduction}
\label{SecIntro}

Heavy fermion (HF) and related intermediate valence (IV) behaviors are
ubiquitous in metallic $f$-electron systems containing lanthanide or actinide
($\equiv M$) atoms with unstable valence.\cite{HFIV}  The HF materials are
typically intermetallic compounds containing Ce, Yb or U ions and are
characterized at the lowest  temperatures $T$ by a large and nearly
$T$-independent spin susceptibility $\chi(T \rightarrow 0) \sim
10^{-2}$\,cm$^3$/(mol~$M$), and an extraordinarly large nearly
$T$-independent electronic specific heat coefficient $\gamma(T \rightarrow 0)
\sim 1$\,J/(mol~$M$)\,K$^2$, where  $\gamma (T) \equiv C_{\rm e}(T)/T$ and
$C_{\rm e}(T)$ is the electronic contribution to the measured specific heat at
constant pressure $C_{\rm p}(T)$.  Large quasiparticle effective masses $m^*$ of
$\sim 100$--1000 electron masses $m_{\rm e}$ have been inferred from $\gamma(0)$
for the HF compounds and smaller values for the IV materials.  The normalized
ratio of $\chi(0)$ to $\gamma(0)$, the Sommerfeld--Wilson ratio\cite{Wilson1975}
$R_{\rm W}$, is on the order of unity in HF and IV materials as in conventional
metals, and is given by $R_{\rm W} = \pi^2 k_{\rm B}^2 \chi(0)/3\mu_{\rm eff}^2
\gamma(0)$, where $k_{\rm B}$ is Boltzmann's constant and $\mu_{\rm eff}$ is the
effective magnetic moment of the Fermi liquid quasiparticles.  For
quasiparticles with (effective) spin $S = 1/2$, one obtains
\begin{equation}
R_{\rm W} = \frac{4 \pi^2 k_{\rm B}^2 \chi(0)}{3 g^2 \mu_{\rm
B}^2 \gamma(0)}~~,
\label{EqRW}
\end{equation} where $g$ is the $g$-factor of the quasiparticles and $\mu_{\rm
B}$ is the Bohr magneton.  Since $R_{\rm W} \sim 1$ in many of the HF and IV
compounds, $\chi$ and $C_{\rm e}$ at low temperatures are both probing the same
low-energy heavy quasiparticle spin  excitations.  With increasing $T$ in the
heaviest-mass systems, $\chi(T)$ crosses over to local-moment behavior and
$\gamma$ decreases rapidly, on a temperature scale of $\sim 0.3$--30\,K\@.

Heavy fermion behaviors are not expected for $d$-electron compounds because of
the much larger spatial extent of $d$ orbitals than of $f$ orbitals and the
resulting stronger hybridization with conduction electron states.  Recently,
however, in collaboration with other researchers, we have documented HF
behaviors, characteristic of those of the heaviest mass $f$-electron HF systems,
in the metallic\cite{Rogers1967} transition metal oxide compound LiV$_2$O$_4$
using $C_{\rm p}(T)$,\cite{Kondo1997}
$\chi(T)$,\cite{Kondo1997,Kondo1998} $^7$Li and $^{51}$V
NMR,\cite{Kondo1997,Mahajan1998} muon spin relaxation
($\mu$SR),\cite{Kondo1997,Merrin1998} and 4--295\,K
crystallography\cite{Kondo1997,Kondo1998,Chmaissem1997} measurements. 
Independent crystallography and $\chi(T)$ measurements\cite{Ueda1997,Onoda1997}
and NMR measurements\cite{Onoda1997,Fujiwara1997,Fujiwara1998} were reported
nearly simultaneously by other groups, with similar results.  LiV$_2$O$_4$ has
the face-centered-cubic normal-spinel structure (space group $Fd{\bar
3}m$),\cite{Reuter1960} and is formally a $d^{1.5}$ system.  The Li atoms occupy
tetrahedral holes and the V atoms octahedral holes in a nearly
cubic-close-packed oxygen sublattice, designated as Li[V$_2$]O$_4$.  The $C_{\rm
e}(T)$ is extraordinarily large for a transition metal compound, $\gamma(1\,{\rm
K})\approx 0.42$\,J/mol\,K$^2$, decreasing rapidly with $T$ to $\sim
0.1$\,J/mol\,K$^2$ at 30\,K.\cite{Kondo1997}  As discussed extensively in
Refs.~\onlinecite{Kondo1997} and~\onlinecite{Kondo1998}, from $\sim 50$--100\,K
to 400\,K, $\chi(T)$ shows a Curie-Weiss-like [$\chi = C/(T - \theta)$] behavior
corresponding to antiferromagnetically coupled ($\theta = -30$ to $-60$\,K)
vanadium local magnetic moments with $S = 1/2$ and $g\approx 2$, but static
magnetic ordering does not occur above 0.02\,K in magnetically pure
LiV$_2$O$_4$, and superconductivity is not observed above 0.01\,K\@.

To our knowledge, in addition to LiV$_2$O$_4$ the only other stoichiometric
transition metal spinel-structure oxide which is metallic to low temperatures
is the normal-spinel compound
LiTi$_2$O$_4$.\cite{Johnston1976,Deschanvres1971,Cava1984,Dalton1994}  In
contrast to LiV$_2$O$_4$, this compound becomes superconducting at $T_{\rm c}
\leq 13.7$\,K (Refs.~\onlinecite{Johnston1976,Johnston1973}) and has a
comparatively $T$-independent and small 
$\chi(T)$ from $T_{\rm c}$ up to 300\,K.
\cite{Johnston1976,Harrison1985,Heintz1989,Tunstall1994}   The resistivity of
thin films at 15\,K is (4.3--8.8)$\times
10^{-4}\,\Omega$\,cm.\cite{Inukai1982}   The spinel system
Li$_{1+x}$Ti$_{2-x}$O$_4$ with cation occupancy Li[Li$_x$Ti$_{2-x}$]O$_4$ exists
from $x = 0$ to
$x = 1/3$;\cite{Johnston1976,Deschanvres1971,Dalton1994} for $x =
1/3$,\cite{Bertaut1953} the oxidation state of the Ti is $+4$ and the compound
is a nonmagnetic insulator.  A zero-temperature superconductor-insulator
transition occurs at $x \sim
0.1$--0.2.\cite{Johnston1976,Harrison1985,Heintz1989}

In this paper, we report the details of our $C_{\rm p}(T)$ measurements on
LiV$_2$O$_4$ and of the data analysis and theoretical modeling.  We have now
obtained data to 108\,K, which significantly extends our previous
high-temperature limit of 78\,K.\cite{Kondo1997}  We also present complementary
linear thermal expansion $\alpha(T)$ measurements on this compound from 4.4 to
297\,K\@.  We will assume that $C_{\rm p}(T)$ can be separated into the sum of
electronic and lattice contributions,
\begin{mathletters}
\label{EqCsum:all}
\begin{equation} C_{\rm p}(T) = C_{\rm e}(T) + C^{\rm lat}(T)~~,
\label{EqCsum:a}
\end{equation}
\begin{equation} C_{\rm e}(T) \equiv \gamma(T)\,T~~.
\label{EqCsum:b}
\end{equation}
\end{mathletters} In Ref.~\onlinecite{Kondo1997}, we reported $C_{\rm p}(T)$
measurements up to 108\,K on Li$_{4/3}$Ti$_{5/3}$O$_4$ which were used to
estimate
$C^{\rm lat}(T)$ in LiV$_2$O$_4$ so that $C_{\rm e}(T)$ could be extracted
according to Eq.~(\ref{EqCsum:a}). In the present work, we report $C_{\rm p}(T)$
up to 108\,K for LiTi$_2$O$_4$, compare these data with those for
Li$_{4/3}$Ti$_{5/3}$O$_4$, and obtain therefrom what we believe to be a more
reliable estimate of $C^{\rm lat}(T)$ for LiV$_2$O$_4$.  The experimental
details are given in Sec.~\ref{SecExpDet}.  An overview of  our $C_{\rm p}(T)$
data for LiV$_2$O$_4$, LiTi$_2$O$_4$ and Li$_{4/3}$Ti$_{5/3}$O$_4$ is given in
Sec.~\ref{SecOverview}.  Detailed analyses of the data for the
Li$_{1+x}$Ti$_{2-x}$O$_4$ compounds and comparisons with literature data are
given in Sec.~\ref{SecLiTiO}, in which we also estimate $C^{\rm lat}(T)$ for
LiV$_2$O$_4$.  The $C_{\rm e}(T)$ and electronic entropy $S_{\rm e}(T)$ for
LiV$_2$O$_4$ are derived in Sec.~\ref{SecLiVO}.  The $\alpha(T)$ measurements are
presented in Sec.~\ref{SecThermExp} and compared with the $C_{\rm p}(T)$ results
and lattice parameter data versus temperature obtained from neutron diffraction
measurements by Chmaissem {\it et al.}\cite{Chmaissem1997}  From the combined
$\alpha(T)$ and $C_{\rm p}(T)$ measurements on the same sample, we derive the
Gr\"uneisen parameter from 4.4 to 108\,K and estimate 
\widetext
\begin{table}
\caption{Lattice parameter $a_0$ and structural [$f_{\rm imp}$ (Str)] and
magnetic [$f_{\rm imp}$ (Mag)] impurity concentrations for the LiV$_2$O$_4$
samples studied in this work.\protect\cite{Kondo1998}}
\begin{tabular}{ccccccc} Sample No.& 2 & 3 & 4A & 5 & 6 \\ \hline   Lattice
parameter (\AA) & 8.23997(4) & 8.24100(15) & 8.24705(29) & 8.24347(25) &
8.23854(11)\\ Impurity phase& V$_2$O$_3$ & pure & V$_2$O$_3$ & V$_2$O$_3$ &
V$_3$O$_5$ \\ 
$f_{\rm imp}$ (Str) (mol\,\%) & 1.83 & $< 1$ & 1.71 & $< 1$ & 2.20 \\
$f_{\rm imp}$ (Mag) (mol\,\%) & 0.22(1) & 0.118(2) & 0.77(2) & 0.472(8) &
0.0113(6) \\
\end{tabular}
\label{TableLiVO}
\end{table}
\narrowtext
\noindent 
the value at $T = 0$. 
Theoretical modeling of the $C_{\rm e}(T)$ data for LiV$_2$O$_4$ is given in
Sec.~\ref{SecTheory}.   Since the electrical resistivity data for single
crystals of LiV$_2$O$_4$ indicate metallic behavior from 4\,K to
450\,K,\cite{Rogers1967} we first discuss the Fermi liquid description of this
compound and derive the effective mass and other parameters for the current
carriers at low temperatures in Sec.~\ref{SecFL}. This is followed by a more
general discussion of the FL theory and its application to LiV$_2$O$_4$ at low
$T$.  In Sec.~\ref{SecMillis} we compare the predictions of Z\"ulicke and
Millis\cite{Zulicke1995} for a quantum-disordered antiferromagnetically coupled
metal with our $C_{\rm e}(T)$ results for LiV$_2$O$_4$.  The isolated $S = 1/2$
impurity Kondo model predicts FL behavior at low temperatures and impurity local
moment behavior at high temperatures.  Precise predictions for the $\chi(T)$ and
$C_{\rm e}(T)$ have been made for this model, and we compare our $C_{\rm e}(T)$
data with those predictions in Sec.~\ref{SecS1/2Kondo}.  In Sec.~\ref{SecHTS} we
consider a local moment model in which the magnetic specific heat of the $B$
sublattice of the $A$[$B_2$]O$_4$ spinel structure for spins $S = 1/2$ and $S =
1$ per $B$ ion is given by a high-temperature series expansion and the
predictions compared with the $C_{\rm e}(T)$ data for LiV$_2$O$_4$.  A summary
and concluding remarks are given in Sec.~\ref{SecConcl}.  Unless otherwise
noted, a ``mol'' refers to a mole of formula units.

\section{Experimental Details}
\label{SecExpDet}

Polycrystalline LiV$_2$O$_4$ samples were prepared using conventional ceramic
techniques described in detail elsewhere, where detailed sample
characterizations and magnetic susceptibility results and analyses are also
given.\cite{Kondo1998}  A few of these results relevant to the present
measurements, analyses and modeling are given in Table~\ref{TableLiVO}.

Polycrystalline LiTi$_2$O$_4$ and Li$_{4/3}$Ti$_{5/3}$O$_4$ samples were
synthesized using solid-state reaction techniques.\cite{Johnston1976}  TiO$_2$
(Johnson Matthey, 99.99\,\%) was dried under a pure oxygen stream at
900\,$^\circ$C before use.  This was mixed with Li$_2$CO$_3$ (Alfa, 99.999\,\%)
in an appropriate ratio to produce either Li$_{4/3}$Ti$_{5/3}$O$_4$ or a
precursor ``LiTiO$_{2.5}$" for LiTi$_2$O$_4$.  The mixtures were then pressed
into pellets and heated at 670\,$^\circ$C in an oxygen atomosphere for $\approx$
1 day.  The weight loss due to release of CO$_2$ was within 0.04 wt.\% of the
theoretical value for LiTiO$_{2.5}$.  However, for Li$_{4/3}$Ti$_{5/3}$O$_4$
additional firings at higher temperatures (up to 800\,$^\circ$C), after being
reground and repelletized, were necessary.  LiTi$_2$O$_4$ was prepared by
heating \newpage
\noindent pressed pellets of a ground mixture of the LiTiO$_{2.5}$ precursor
and Ti$_2$O$_3$ in an evacuated and sealed quartz tube at 700\,$^\circ$C for one
week and then air-cooling.  The Ti$_2$O$_3$ was prepared by heating a mixture of
TiO$_2$ and titanium metal powder (Johnson Matthey) at 1000\,$^\circ$C for one
week in an evacuated and sealed quartz tube. 

Powder x-ray diffraction data were obtained using a Rigaku diffractometer (Cu
K$\alpha$ radiation) with a curved graphite crystal monochromator.  Rietveld
refinements of the data were carried out using the  program ``Rietan 97 (`beta'
version)''.\cite{Izumi1993}  The x-ray data for our sample of
Li$_{4/3}$Ti$_{5/3}$O$_4$ showed a  nearly pure spinel phase with a trace of
TiO$_2$ (rutile) impurity phase.  The two-phase refinement, assuming the cation
distribution Li[Li$_{1/3}$Ti$_{5/3}$]O$_4$, yielded the lattice
$a_0$ and oxygen $u$ parameters of the spinel phase  8.3589(3)\,\AA\ and
0.2625(3), respectively; the concentration of TiO$_2$ impurity phase was
determined to be 1.3\,mol\%.  The LiTi$_2$O$_4$ sample 
 was nearly a single-phase
spinel structure but with a trace of Ti$_2$O$_3$ impurity.  A two-phase Rietveld
refinement assuming the normal-spinel cation distribution yielded the spinel
phase parameters
$a_0$ = 8.4033(4)\,\AA\ and $u$ = 0.2628(8), and the Ti$_2$O$_3$ impurity phase
concentration $<1$\,mol\%.  Our crystal data are compared with those of
Cava~{\it et al.}\cite{Cava1984} and Dalton~{\it et al.}\cite{Dalton1994} in
Table~\ref{TableI}.

The $C_{\rm p}(T)$ measurements were done on samples from four different batches
of LiV$_2$O$_4$ using a conventional heat-pulse calorimeter, with Apeizon-N
grease providing contact between the sample and the copper
tray.\cite{Swenson1996}  Additional $C_{\rm p}(T)$ data were obtained up to
108\,K on 0.88\,g of the isostructural nonmagnetic insulator spinel compound
Li$_{4/3}$Ti$_{5/3}$O$_4$, containing only maximally oxidized Ti$^{+4}$, and
3.09\,g of the isostructural superconductor LiTi$_2$O$_4$ to obtain an estimate
of the background lattice contribution.  A basic limitation on the accuracy of
these $C_{\rm p}$ data, except for LiV$_2$O$_4$ below 15\,K, was the relatively
small (and sample-dependent) ratios of the heat capacities of the samples to
those associated with the tray (the 
\widetext
\begin{table}
\caption{Characteristics of LiTi$_2$O$_4$ and Li$_{4/3}$Ti$_{5/3}$O$_4$
samples.  Abbreviations:
$a_0$ is the lattice parameter, $u$ the oxygen parameter, $\gamma$ the
electronic specific heat coefficient, $\theta_0$ the zero-temperature Debye
temperature, $T_{\rm c}$ and $\Delta T_{\rm c}$ the superconducting transition
temperature and transition width, and $\Delta C_{\rm p}$ the specific heat jump
at $T_{\rm c}$.}
\begin{tabular}{lldcdddr}
$a_0$ & $u$ & $\gamma$ & $\theta_0$ & $T_{\rm c}$ & $\Delta T_{\rm c}$ & $\Delta
C_{\rm p}/\gamma T_{\rm c}$ & Ref.\\  (\AA)  &  & (mJ/mol\,K$^2$)  & (K) & (K) 
&  (K)  &  (mJ/mol\,K$^2$)  &  \\
\hline &&&LiTi$_2$O$_4$\\
\hline 8.4033(4) & 0.2628(8) & 17.9(2) & 700(20) & 11.8 & $\lesssim$0.2 &
1.75(3) & This Work\\ 8.4033(1) & 0.26275(5) &&&&&&\onlinecite{Cava1984}\\
8.41134(1) & 0.26260(4) &&&&&&\onlinecite{Dalton1994}\\ 8.407 &  & 21.4 & 685 &
11.7 & 1.2 & 1.59 & \onlinecite{McCallum1976}\\ & & 22.0 & 535 & 12.4  & 0.32 
&  1.57  &\onlinecite{Heintz1989}\\ & 0.26290(6)
(300\,K)&&&&&&\onlinecite{Tunstall1994}\\ & 0.26261(5)
(6\,K)&&&&&&\onlinecite{Tunstall1994}\\
\hline &&&Li$_{4/3}$Ti$_{5/3}$O$_4$\\
\hline 8.3589(3) & 0.2625(3) & 0. & 725(20) &&&&This Work\\ 8.35685(2) &
0.26263(3) &&&&&&\onlinecite{Dalton1994}\\ 8.359 & & 0. & 610
&&&&\onlinecite{McCallum1976}\\ & & 0.05 & 518 &&&&\onlinecite{Heintz1989}\\
\end{tabular}
\label{TableI}
\end{table}
\narrowtext
\noindent
``addenda'').  For LiV$_2$O$_4$ sample 6,
this ratio decreased from 40 near 1\,K to 1.0 at 15\,K to a relatively constant
0.2 above 40\,K\@.  For the superconducting LiTi$_2$O$_4$ sample, this ratio was
0.45 just above $T_{\rm c}$ ($= 11.8$\,K), and increased to 0.65 at 108\,K\@. 
For the nonmagnetic insulator Li$_{4/3}$Ti$_{5/3}$O$_4$ sample, this ratio
varied from 0.03 to 0.12 to 0.2 at 8, 20 and 108\,K, respectively.  These
factors are important since small ($\pm 0.5$\%) systematic uncertainties in the
addenda heat capacity can have differing effects on the $C_{\rm p}(T)$ measured
for the different samples, even though the precision of the raw heat capacity
measurements (as determined from fits to the data) is better than 0.25\%.

The linear thermal expansion coefficient of LiV$_2$O$_4$ sample~6  was measured
using a differential capacitance dilatometer.\cite{Swenson1996,Swenson1998}  All
data were taken isothermally ($T$ constant to 0.001\,K).  The absolute accuracy
of the measurements is estimated to be better than 1\%.

\section{Specific Heat Measurements}
\label{SecExpRes}

\subsection{Overview}
\label{SecOverview}

\begin{figure}
\epsfxsize=3.1 in
\epsfbox{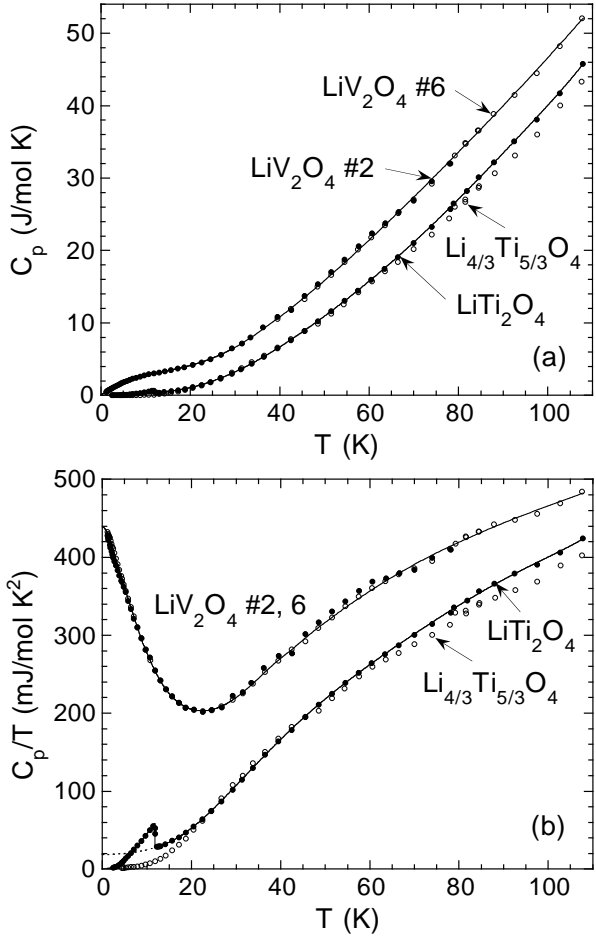}
\vglue 0.1in
\caption{Overview of the molar specific heat $C_{\rm p}$ (a) and $C_{\rm p}/T$
(b) {\it vs.}\ temperature $T$ for LiV$_2$O$_4$ samples~2 ($\bullet$) and~6
($\circ$) and the reference compounds LiTi$_2$O$_4$ ($\bullet$), a metallic
superconductor, and Li$_{4/3}$Ti$_{5/3}$O$_4$ ($\circ$), a nonmagnetic
insulator.  The solid curves are polynomial fits to the data for LiV$_2$O$_4$
sample~6 and LiTi$_2$O$_4$.  The dashed curve in (b) is the inferred normal
state $C_{\rm p}/T$ below $T_{\rm c}$ for LiTi$_2$O$_4$.\label{FigCpSumm}}
\end{figure}

An overview of our $C_{\rm p}(T)$ measurements on LiV$_2$O$_4$ sample 2, run 2
(1.26--78\,K), sample~6 (1.16--108\,K), and LiTi$_2$O$_4$ and
Li$_{4/3}$Ti$_{5/3}$O$_4$ up to 108\,K, is shown in plots of $C_{\rm p}(T)$ and
$C_{\rm p}(T)/T$ in Figs.~\ref{FigCpSumm}(a) and~(b), respectively.  Our data
for LiTi$_2$O$_4$ and Li$_{4/3}$Ti$_{5/3}$O$_4$ are generally in agreement with
those of McCallum {\em et al.}\cite{McCallum1976} which cover the range up to
$\sim 25$\,K\@.  For LiTi$_2$O$_4$ above $T_{\rm c} = 11.8$\,K (see below) and
for Li$_{4/3}$Ti$_{5/3}$O$_4$, one sees from Fig.~\ref{FigCpSumm}(a) a smooth
monotonic increase in $C_{\rm p}$ up to 108\,K\@.  From Fig.~\ref{FigCpSumm}(b),
the $C_{\rm p}$ of the  nonmagnetic insulator Li$_{4/3}$Ti$_{5/3}$O$_4$ is
smaller than that of metallic LiTi$_2$O$_4$ up to $\sim 25$\,K, is larger up to
$\sim 45$\,K and then becomes smaller again at higher temperatures.  Since
$C_{\rm e} = 0$ in Li$_{4/3}$Ti$_{5/3}$O$_4$ and $C_{\rm e}(T)$ in LiTi$_2$O$_4$
cannot be negative,
\noindent
it follows from Eq.~(\ref{EqCsum:a}) that $C^{\rm
lat}(T)$ and hence the lattice dynamics are significantly different in
LiTi$_2$O$_4$ compared with Li$_{4/3}$Ti$_{5/3}$O$_4$.  The data for LiV$_2$O$_4$
in Fig.~\ref{FigCpSumm}(b) are shifted upwards from the data for the Ti spinels,
with  a strong upturn in $C_{\rm p}(T)/T$ below $\sim 25$\,K\@.  These data
indicate a very large
$\gamma(T\rightarrow 0)$.  Comparison of $C_{\rm p}(T)/T$ for LiV$_2$O$_4$ and
LiTi$_2$O$_4$ at the higher temperatures $> 30$\,K indicate that a large
$\gamma(T)$ persists in LiV$_2$O$_4$ up to our maximum measurement temperature
of 108\,K\@.  In the following, we begin our  analyses with the data for the
Li$_{1+x}$Ti$_{2-x}$O$_4$ compounds because we extract a lattice specific heat
from these materials as a reference for LiV$_2$O$_4$.

\subsection{Li$_{1+x}$Ti$_{2-x}$O$_4$}
\label{SecLiTiO}

In the present paper, our $C_{\rm p}(T)$ data for LiTi$_2$O$_4$ and
Li$_{4/3}$Ti$_{5/3}$O$_4$ are most important for determining the lattice
contribution $C^{\rm lat}(T)$ to $C_{\rm p}(T)$ of LiV$_2$O$_4$.  At low
temperatures, the $C_{\rm p}(T)$ of a conventional nonmagnetic,
nonsuperconducting material is\cite{Gopal1966}
\begin{equation} C_{\rm p}(T) = A_1 T + A_3 T^3 + A_5 T^5 + A_7 T^7 + \cdots~~,
\label{EqCLoT}
\end{equation} where $\gamma \equiv A_1$ and $\beta \equiv A_3$.  From
Eqs.~(\ref{EqCsum:all}), the first term in Eq.~(\ref{EqCLoT}) is $C_{\rm e}(T)$,
the second corresponds to the ideal Debye lattice contribution $C^{\rm
lat}(T\rightarrow 0)$ and the following terms represent dispersion in the
lattice properties.\cite{Barron1980}  The  zero-temperature Debye temperature
$\theta_0$ is given by\cite{Gopal1966} $\theta_0 = (1.944 \times 10^6
r/\beta)^{1/3}$, where $r$ is the number of atoms per formula unit ($r = 7$
here) and $\beta$ is in units of mJ/mol\,K$^4$.  Equation~(\ref{EqCLoT})
suggests the commonly used plot of $C_{\rm p}/T$ versus
$T^2$ to obtain the parameters $\gamma$ and $\beta$.  Unfortunately, the very
small heat capacity of the small Li$_{4/3}$Ti$_{5/3}$O$_4$ sample and the
occurrence of the superconducting transition in LiTi$_2$O$_4$ at 11.8\,K
complicate the use of this relation to determine $C^{\rm lat}(T)$ for these
presumably similar materials below $\approx 12$\,K\@.

\begin{figure}
\epsfxsize=3.3 in
\epsfbox{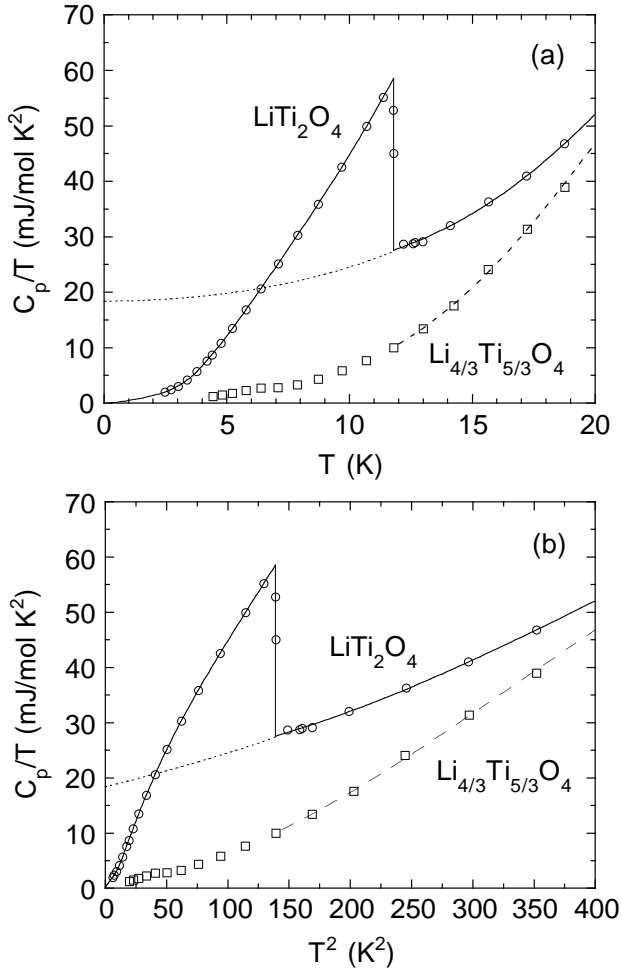}
\vglue 0.1in
\caption{Expanded plots below 20\,K of the molar specific heat divided by
temperature $C_{\rm p}/T$ {\it vs.}\ temperature $T$ of LiTi$_2$O$_4$ and
Li$_{4/3}$Ti$_{5/3}$O$_4$ from Fig.~\protect\ref{FigCpSumm}.  The solid curves
are polynomical fits to the data for LiTi$_2$O$_4$, whereas the dotted curve is
the inferred normal state behavior below $T_{\rm c} = 11.8$\,K\@.  The dashed 
curve is a polynomial fit to the data for Li$_{4/3}$Ti$_{5/3}$O$_4$ 
above 12\,K\@.\label{FigC/TLi1}}
\end{figure}

The $C_{\rm p}(T)/T$ of LiTi$_2$O$_4$ below 20\,K is plotted versus $T$ and
$T^2$ in Fig.~\ref{FigC/TLi1}(a) and (b), respectively.  The superconducting
transition at $T_{\rm c} = 11.8$\,K is seen to be pronounced and very sharp
($\Delta T_{\rm c} \lesssim 0.2$\,K).  The dotted line extrapolation of the
normal state ($T > 11.8$\,K) data to $T = 0$ shown in Figs.~\ref{FigCpSumm}(b)
and~\ref{FigC/TLi1} uses Eq.~(\ref{EqCLoT}), equality of the superconducting and
normal state entropy at $T_{\rm c}$, $S(11.8\,{\rm K}) = 241(1)$\,mJ/mol\,K, and
continuity considerations with $C_{\rm p}(T)/T$ above $T_{\rm c}$, from which we
also obtain estimates of $\gamma$ and $\beta$.  Although we cannot rule out a
$T$-dependence of $\gamma$, we assume here that $\gamma$ is independent of $T$. 
While $\gamma$ [$= 17.9$(2)\,mJ/mol\,K$^2$] appears to be quite insensitive to
addenda uncertainties, $\theta_0$ [$= 700$(20)\,K] is less well-defined.  Our
value for $\gamma$ is slightly smaller than the values of 20--22\,mJ/mol\,K$^2$
reported earlier for LiTi$_2$O$_4$,\cite{Heintz1989,McCallum1976} as shown in
Table~\ref{TableI}.  From the measured superconducting state $C_{\rm p}(T_{\rm
c}) = 684(2)$\,mJ/mol\,K and normal state $C_{\rm p}(T_{\rm c}) =
315(1)$\,mJ/mol\,K, the discontinuity in $C_{\rm p}$ at $T_{\rm c}$ is given by
$\Delta C_{\rm p}/T_{\rm c} = 31.3(3)$\,mJ/mol\,K$^2$, yielding $\Delta C_{\rm
p}/\gamma T_{\rm c} = 1.75(3)$ which is slightly larger than previous estimates
in Table~\ref{TableI}.  According to Eqs.~(\ref{EqCsum:all}), the lattice
specific heat of LiTi$_2$O$_4$ above $T_{\rm c}$ is given by $C^{\rm lat}(T) =
C_{\rm p}(T) - \gamma T$.

The $C^{\rm lat}(T)$ derived for LiTi$_2$O$_4$ below 12\,K is consistent within
experimental uncertainties with the measured $C^{\rm lat}(T)$ of
Li$_{4/3}$Ti$_{5/3}$O$_4$ in the same temperature range after accounting for the
formula weight difference.  The low-$T$ $C_{\rm p}(T)/T = C^{\rm lat}(T)/T$ for
Li$_{4/3}$Ti$_{5/3}$O$_4$ is plotted in Figs.~\ref{FigC/TLi1}.  The $\theta_0 =
725$(20)\,K found for Li$_{4/3}$Ti$_{5/3}$O$_4$ is slightly larger than that for
LiTi$_2$O$_4$, as expected.  A polynomial fit to the $C_{\rm p}(T)$ of
Li$_{4/3}$Ti$_{5/3}$O$_4$ above 12\,K is shown by the dashed curves in
Figs.~\ref{FigC/TLi1}.  The uncertainties in the data and analyses for the Ti
spinels have little effect on the analyses of $C_{\rm p}(T)$ for LiV$_2$O$_4$ in
the following Sec.~\ref{SecLiVO}, since as Figs.~\ref{FigCpSumm} suggest,
$C^{\rm lat}(T)$ for LiV$_2$O$_4$ is small compared to $C_{\rm e}(T)$ of this
compound at low temperatures.  

\begin{figure}
\epsfxsize=3.1 in
\epsfbox{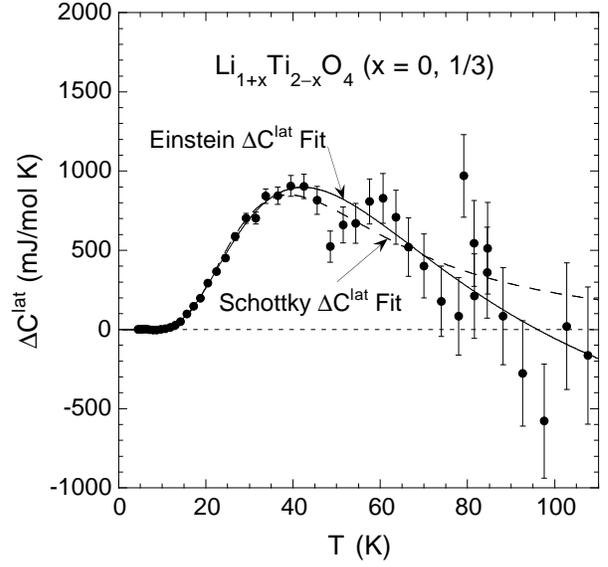}
\vglue 0.1in
\caption{The difference $\Delta C^{\rm lat}$ between the lattice specific heats
of Li$_{4/3}$Ti$_{5/3}$O$_4$ and LiTi$_2$O$_4$  {\it vs}.\ temperature $T$.  The
solid curve is a fit to the data by the difference between two Einstein specific
heats in Eq.~(\protect\ref{EqDClattFit}), whereas the dashed curve is the
Schottky specific heat of a two-level system in
Eq.~(\protect\ref{EqDClattSchFit}).  The error bars represent $\pm$1\% of the
measured $C_{\rm p}(T)$ for Li$_{4/3}$Ti$_{5/3}$O$_4$.\label{FigDClatt}}
\end{figure}

To quantify the difference above $\sim 12$\,K between the $C^{\rm lat}(T)$ of
LiTi$_2$O$_4$ and Li$_{4/3}$Ti$_{5/3}$O$_4$ noted above in
\noindent 
Sec.~\ref{SecOverview}, in Fig.~\ref{FigDClatt} is plotted the difference
$\Delta C^{\rm lat}(T)$ between the measured $C_{\rm p}(T)$ of
Li$_{4/3}$Ti$_{5/3}$O$_4$ and $C^{\rm lat}(T)$ of LiTi$_2$O$_4$.  The shape of
$\Delta C^{\rm lat}(T)$ in Fig.~\ref{FigDClatt} below $\sim 30\,K$ is similar to
that of an Einstein specific heat, but such a specific heat saturates to the
Dulong-Petit limit at high $T$ and does not decrease with $T$ as the data do
above 40\,K\@.  These observations suggest that intermediate-energy phonon modes
in LiTi$_2$O$_4$ at some energy $k_{\rm B} T_{\rm E2}$ split in
Li$_{4/3}$Ti$_{5/3}$O$_4$ into higher ($k_{\rm B} T_{\rm E3}$) and lower
($k_{\rm B} T_{\rm E1}$) energy modes, resulting from the Li-Ti atomic disorder
on the octahedral sites in Li$_{4/3}$Ti$_{5/3}$O$_4$ and/or from the difference
in the metallic character of the two compounds.  Following this interpretation,
we model the data in Fig.~\ref{FigDClatt} as the difference $\Delta C^{\rm
lat}_{\rm Einstein}$ between the Einstein heat capacities of two Einstein modes
with Einstein temperatures of $T_{\rm E1}$ and
$T_{\rm E2}$ (neglecting the modes at high energy $k_{\rm B} T_{\rm E3}$), given
by\cite{Gopal1966}
\begin{equation}
\Delta C^{\rm lat}_{\rm Einstein} = 3rR\Bigg[\frac{x_1(T_{\rm
E1}/2T)^2}{\sinh^2(T_{\rm E1}/2T)} - \frac{x_2(T_{\rm E2}/2T)^2}{\sinh^2(T_{\rm
E2}/2T)}\Bigg]~~,
\label{EqDClattFit}
\end{equation} where $R$ is the molar gas constant, $r$ = 7 atoms/formula unit
and $x_1$ and $x_2$ are the fractions of the total number of phonon modes
shifted to $T_{\rm E1}$ and away from $T_{\rm E2}$, respectively.  A  reasonable
fit of the data by Eq.~(\ref{EqDClattFit}) was obtained with the parameters
$x_1$ = 0.012, $T_{\rm E1} = 110$\,K, $x_2 = 0.018$ and $T_{\rm E2} = 240$\,K;
the fit is shown as the solid curve in Fig.~\ref{FigDClatt}.  The model then
predicts that a fraction $(x_2 - x_1)/x_2 \sim 0.3$ of the modes removed at
energy
$k_{\rm B} T_{\rm E2}$ are moved to an energy $k_{\rm B} T_{\rm E3} \gg k_{\rm
B} T_{\rm E2}$.
 
An alternative parametrization of the experimental $\Delta C^{\rm lat}(T)$ data
can be given in terms of the specific heat of a two-level system, described by
the Schottky function\cite{Gopal1966}
\begin{equation}
\Delta C^{\rm lat}_{\rm Schottky} = xrR\bigg({g_0\over
g_1}\bigg)\bigg({\delta\over T}\bigg)^2\frac{{\rm e}^{\delta/T}}{\big[1 +
(g_0/g_1){\rm e}^{\delta/T}\big]^2}~~,
\label{EqDClattSchFit}
\end{equation} where $x$ is the atomic fraction of two-level sites, $g_0$ and
$g_1$ are respectively the degeneracies of the ground and excited levels and
$\delta$ is the energy level splitting in temperature units.  Fitting
Eq.~(\ref{EqDClattSchFit}) to the data in Fig.~\ref{FigDClatt}, we find $g_1/g_0
= 4$, $x = 0.012$ and $\delta = 117$\,K\@.  The fit is shown as the dashed curve
in Fig.~\ref{FigDClatt}.  The accuracy of our $\Delta C^{\rm lat}(T)$ data is
not sufficient to discriminate between the applicability of the Einstein and
Schottky descriptions.

\subsection{LiV$_2$O$_4$}
\label{SecLiVO}

Specific heat $C_{\rm p}(T)$ data were obtained for samples from four batches of
LiV$_2$O$_4$.  Our first experiment was carried out on sample~2 (run~1) with
mass 5~g.  The $C_{\rm p}(T)$ was found to be so large at low $T$ (the first
indication of heavy fermion behavior in this compound from these measurements)
that the large thermal diffusivity limited our measurements to the 2.23--7.94\,K
temperature range.  A smaller piece of sample~2 (0.48~g) was then measured
(run~2) from 1.16 to 78.1\,K\@.  Data for samples from two additional batches
(sample~3 of mass 0.63~g, 1.17--29.3\,K, and sample~4A of mass 0.49~g,
1.16--39.5\,K) were also obtained.  Subsequent to the theoretical modeling of
the data for sample~3 described below in Sec.~\ref{SecTheory}, we obtained a
complete data set from 1.14 to 108\,K for sample~6 with mass 1.1\,g from a
fourth batch.  A power series fit to the $C_{\rm p}(T)$ data for sample~6 is
shown as solid curves in Fig.~\ref{FigCpSumm}.

We have seen above that $C^{\rm lat}(T)$ of LiTi$_2$O$_4$ is significantly
different from that of Li$_{4/3}$Ti$_{5/3}$O$_4$.  Since LiV$_2$O$_4$ is a
metallic normal-spinel compound with cation occupancies Li[V$_2$]O$_4$ as in
Li[Ti$_2$]O$_4$, and since the formula weight of metallic  LiTi$_2$O$_4$ is much
closer to that of LiV$_2$O$_4$ than is that of the insulator
Li$_{4/3}$Ti$_{5/3}$O$_4$, we expect that the lattice dynamics and
$C^{\rm lat}(T)$ of LiV$_2$O$_4$ are much better approximated by those of
LiTi$_2$O$_4$ than of Li$_{4/3}$Ti$_{5/3}$O$_4$.  Additionally, more precise and
accurate $C_{\rm p}(T)$ data were obtained for LiTi$_2$O$_4$ as compared to 
Li$_{4/3}$Ti$_{5/3}$O$_4$ because of the factor of three larger mass of the
former compound measured than of the latter.  Therefore, we will assume in the
following that the $C^{\rm lat}(T)$ of LiV$_2$O$_4$ from 0--108\,K is identical
with that given above for LiTi$_2$O$_4$.  We do not attempt to correct for the
influence of the small formula weight difference of 3.5\% between these two
compounds on $C^{\rm lat}(T)$; this difference would only be expected to shift
the Debye temperature by $\lesssim 1.8$\%, which is on the order of the accuracy
of the high temperature $C_{\rm p}(T)$ data.  The
$C_{\rm e}(T)$ of LiV$_2$O$_4$ is then obtained using Eq.~(\ref{EqCsum:a}).  

\begin{figure}
\epsfxsize=3.3 in
\epsfbox{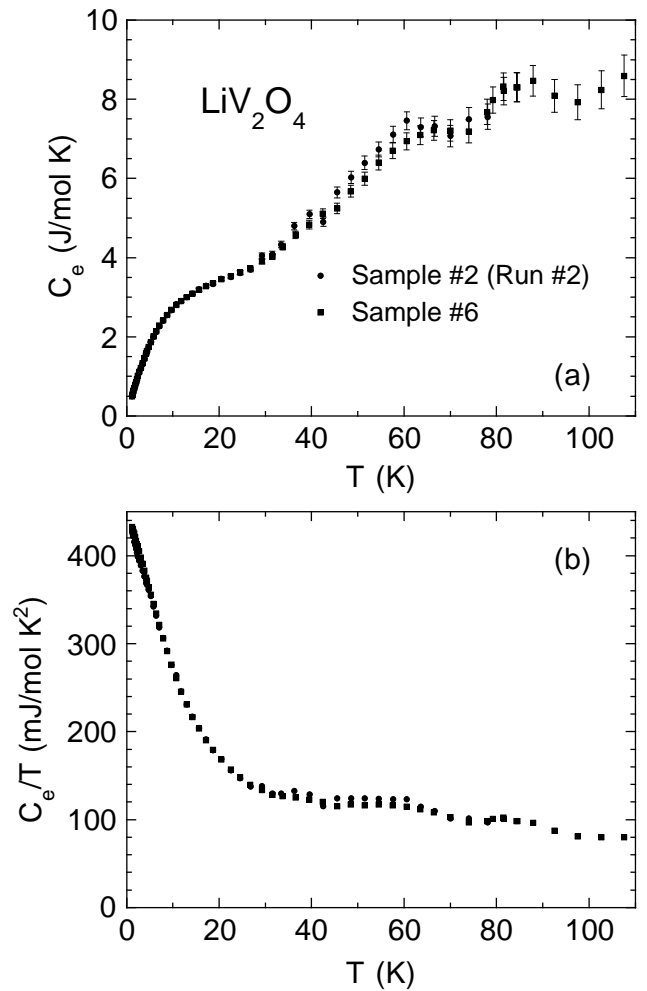}
\vglue 0.1in
\caption{Electronic specific heat $C_{\rm e}$ (a) and $C_{\rm e}/T$ (b)  {\it
vs.}\ temperature $T$ for LiV$_2$O$_4$ samples~2 (run~2) and~6.  The error bars
in (a) represent $\pm$1\% of the measured $C_{\rm p}(T)$ for
LiV$_2$O$_4$.\label{FigCe}}
\end{figure}

\begin{figure}
\epsfxsize=3.1 in
\epsfbox{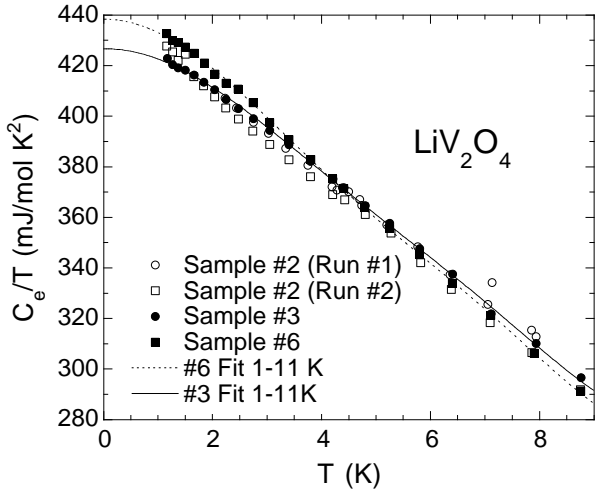}
\vglue 0.1in
\caption{Expanded plot below 9\,K of the $C_{\rm e}/T$ {\it vs.}\ $T$ data for
LiV$_2$O$_4$ samples~2, 3 and~6.  The solid and dashed curves are polynomial
fits to the 1.1--10\,K data for samples~3 and~6, respectively.\label{FigCe2}}
\end{figure}

The $C_{\rm e}(T)$ data for samples~2 (run~2) and~6 of LiV$_2$O$_4$, obtained
using Eqs.~(\ref{EqCsum:all}), are shown up to 108\,K in plots of $C_{\rm e}(T)$
and $C_{\rm e}(T)/T$ {\it vs.}\ $T$ in Figs.~\ref{FigCe}(a) and (b),
respectively.  An expanded plot of $C_{\rm e}(T)$ below 9\,K for LiV$_2$O$_4$ is
shown in Fig.~\ref{FigCe2}(a), where data for sample~2 (run~1) and sample~3 are
also included.  The data for samples~2 and~3 are seen to be in agreement to
within about 1\%.  However, there is a small  positive curvature in the data for
sample~2 below $\sim 3$\,K, contrary to the small negative curvature for
sample~3.  This difference is interpreted to reflect the influence of the larger
magnetic defect concentration present in sample~2 as compared with that in
sample~3, see Table~\ref{TableLiVO}.\cite{Kondo1998}  Therefore, we believe that
the $C_{\rm e}(T)$ data for sample~3 more closely reflect the intrinsic behavior
of defect-free LiV$_2$O$_4$ compared to the data for sample~2 and all fits to
$C_{\rm e}(T)$ of LiV$_2$O$_4$ below 30\,K by theoretical models to be presented
in Sec.~\ref{SecTheory} below are therefore done using the data for sample~3. 
As seen 
\noindent
in  Fig.~\ref{FigCe2}(a), the $C_{\rm e}(T)$ data for sample~6 lie
somewhat higher than the data for the other samples below about 4\,K but are
comparable with those for the other samples at higher temperatures.  This
difference is also reflected in the magnetic susceptibilities
$\chi(T)$,\cite{Kondo1998} where $\chi(T)$ for sample~6 is found to be slightly
larger than those of other samples.

To obtain extrapolations of the electronic specific heat to $T = 0$, the $C_{\rm
e}(T)/T$ data in Fig.~\ref{FigCe2} from 1 to 10\,K for samples~3 and~6 were
fitted by the  polynomial 
\begin{equation} {C_{\rm e}(T)\over T} = \gamma(0) + \sum_{n=1}^5 C_{2n}
T^{2n}~~, 
\label{EqCeFit}
\end{equation} yielding
\begin{mathletters}
\label{EqGam(0):all}
\begin{equation}
\gamma(0) = 426.7(6)\,{\rm mJ/mol\,K^2}~~{\rm (sample\ 3)}~,\label{EqGam(0):a}
\end{equation}
\begin{equation}
\gamma(0) = 438.3(5)\,{\rm mJ/mol\,K^2}~~{\rm (sample\ 6)}~.\label{EqGam(0):b}
\end{equation}
\end{mathletters}

\noindent
The fits for samples~3 and~6 are respectively shown by solid and dashed curves in
Fig.~\ref{FigCe2}. The $\gamma(0)$ values are an order of magnitude or more
larger than typically obtained for transition metal compounds, and are about 23
times larger than found above for LiTi$_2$O$_4$.

\begin{figure}
\epsfxsize=3.1 in
\epsfbox{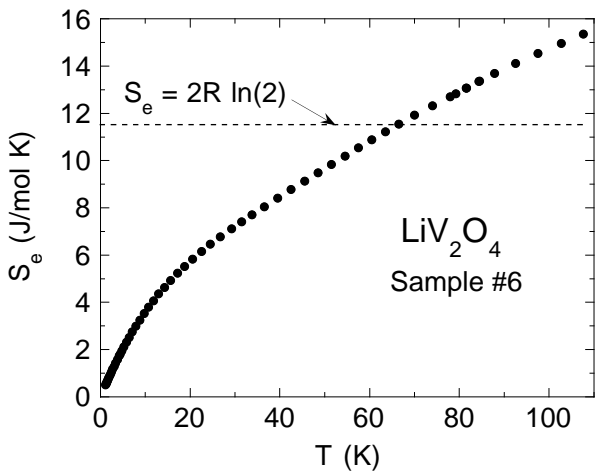}
\vglue 0.1in
\caption{Electronic entropy $S_{\rm e}$ of LiV$_2$O$_4$ sample~6 versus
temperature $T$
 ($\bullet$), obtained by integrating the $C_{\rm e}/T$ data for sample~6 in
Fig.~\protect\ref{FigCe}(b) with $T$.\label{FigSe}}
\end{figure}

The $T$-dependent electronic entropy $S_{\rm e}(T)$ of LiV$_2$O$_4$ was obtained
by integrating the $C_{\rm e}(T)/T$ data for sample 6 in Fig.~\ref{FigCe}(b)
with $T$; the extrapolation of the
$C_{\rm e}(T)/T$ {\it vs.}\ $T$ fit for sample~6 in Fig.~\ref{FigCe2} from $T =
1.16$\,K to $T=0$ yields an additional entropy of $S_{\rm e}($1.16\,K) =
0.505\,J/mol\,K\@.  The total $S_{\rm e}(T)$ is shown up to 108\,K in
Fig.~\ref{FigSe}; these data are nearly identical with those of sample~2 (run~2)
up to the maximum measurement temperature of 78~K for that sample (not shown). 
The electronic entropy at the higher temperatures is large.  For example, if
LiV$_2$O$_4$ were to be considered to be a strictly 
\begin{figure}
\epsfxsize=3.2 in
\epsfbox{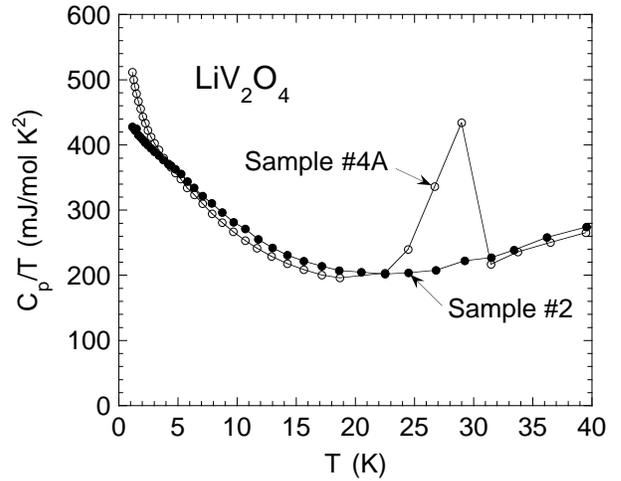}
\vglue 0.1in
\caption{Measured specific heat divided by temperature $C_{\rm p}/T$ {\it vs.}\
temperature $T$ for LiV$_2$O$_4$ sample~4A; corresponding data for sample~2 from
Fig.~\protect\ref{FigCpSumm} are shown for comparison.  The lines are guides to
the eye.\label{FigSam4A}}
\end{figure}
\noindent
localized moment system with
one spin $S = 1/2$ per V atom, then the maximum electronic (spin) entropy would
be 2$R\ln(2)$, which is already reached at  about 65\,K as shown by comparison
of the data with the horizontal dashed line in Fig.~\ref{FigSe}.
Our $C_{\rm p}(T)$ data for one sample (sample~4A) of LiV$_2$O$_4$ were
anomalous.  These are shown in Fig.~\ref{FigSam4A} along with those of sample~2
(run~2) for comparison.  Contrary to the $C_{\rm p}(T)/T$ data for sample~2, the
data for sample~4A show a strong upturn below $\sim 5$\,K and a peak at about
29\,K\@.  We have previously associated the first type of effect with
significant ($\sim 1$\,mol\%) concentrations of paramagnetic
defects.\cite{Kondo1997}  Indeed, Table~\ref{TableLiVO} shows that this sample
has by far the highest magnetic impurity concentration of all the samples we
studied in detail.  The anomalous peak at 29\,K might be inferred to be due to
small amounts of impurity phases (see Table~\ref{TableLiVO}).   However, the
excess entropy $\Delta S$ under the peak is rather large,
$\Delta S \sim 0.9\,{\rm J/mol\,K} \approx 0.16R$ln(2).  We also note that the
height of the anomaly above ``background'' is at least an order of magnitude
larger than would be anticipated due to a few percent of V$_4$O$_7$ or
V$_5$O$_9$ impurity phases which order antiferromagnetically with N\'eel
temperatures of 33.3 and 28.8\,K, respectively.\cite{Khattak1978}  It is
possible that the 29\,K anomaly is intrinsic to the spinel phase in this
particular sample; in such a case Li-V antisite disorder and/or other types of
crystalline defects would evidently be involved.  As seen in
Table~\ref{TableLiVO}, this sample has by far the largest room temperature
lattice parameter of all the samples listed, which may be a reflection of a
slightly different stoichiometry and/or defect distribution or concentration
from the other samples.  Although these $C_{\rm p}(T)$ data for sample~4A will
not be discussed further in this paper, the origin of the anomaly at 29\,K
deserves further investigation.

\begin{figure}
\epsfxsize=3.2 in
\epsfbox{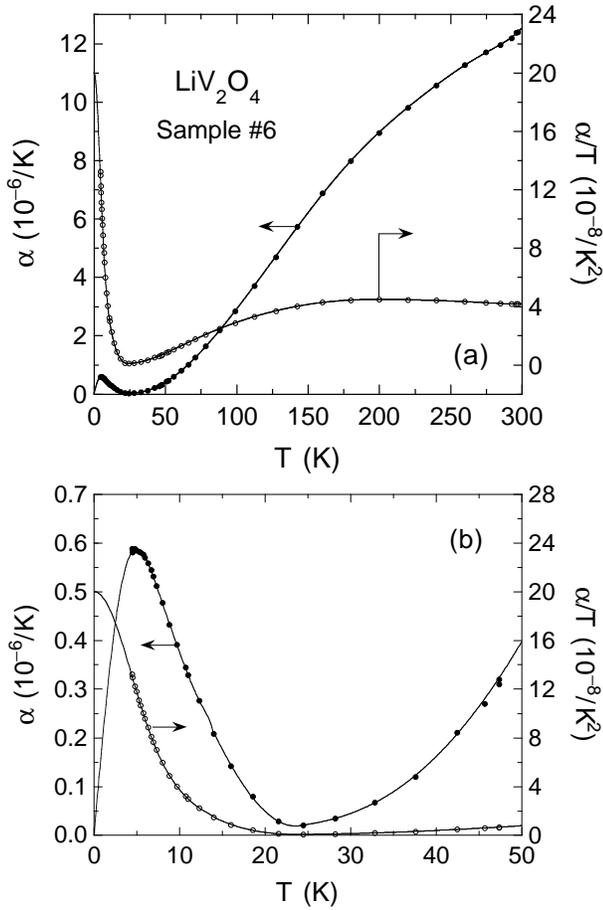}
\vglue 0.1in
\caption{Linear thermal expansion coefficient $\alpha$ (left-hand scales) and
$\alpha/T$ (right-hand scales) versus temperature $T$ for LiV$_2$O$_4$ sample~6
from 4.4 to 297\,K (a) and 4.4 to 50\,K (b).  The solid curves are the fit to the
$\alpha(T)$ data by a polynomial.}
\label{FigAlpha3}
\end{figure}

\section{Thermal Expansion Measurements}
\label{SecThermExp}

The linear thermal expansion coefficient $\alpha(T)$ of LiV$_2$O$_4$ sample~6 was
measured between 4.4 and 297\,K\@.  Figure~\ref{FigAlpha3}(a) shows $\alpha(T)$
and
$\alpha(T)/T$ over this $T$ range, and Fig.~\ref{FigAlpha3}(b) shows expanded
plots below 50\,K\@.  At 297\,K, $\alpha = 12.4 \times 10^{-6}$\,K$^{-1}$, which
may be compared with the value  $\alpha \approx 15.6 \times 10^{-6}$\,K$^{-1}$
obtained for LiTi$_2$O$_4$ between 293 and 1073\,K from x-ray diffraction
measurements.\cite{Roy1977} Upon cooling from 297\,K to about 25\,K, $\alpha$ of
LiV$_2$O$_4$ decreases as is typical of conventional metals.\cite{Barron1980} 
However, $\alpha(T)$ nearly becomes negative with decreasing $T$ at about
23\,K\@.  This trend is preempted upon further cooling below $\sim 20$\,K, where
both $\alpha(T)$ and $\alpha(T)/T$ exhibit strong increases.  The strong
increase in $\alpha(T)$ below 20\,K was first observed by
Chmaissem~{\it et~al}.\cite{Chmaissem1997} from high-resolution neutron
diffraction data, which motivated the present $\alpha(T)$ measurements.  We
fitted our $\alpha(T)$ data by a polynomial in $T$ over three contiguous
temperature ranges and obtained the fit shown as the solid curves in
Figs.~\ref{FigAlpha3}.  From the fit, we obtain 
\begin{figure}
\epsfxsize=3.2 in
\epsfbox{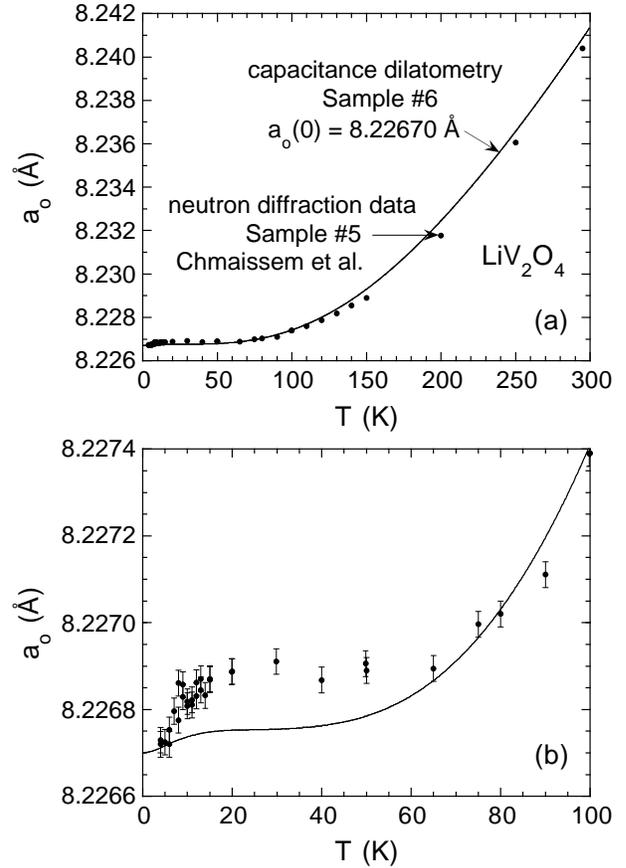}
\vglue 0.1in
\caption{Lattice parameter $a_{\rm o}$ versus temperature $T$ from 4 to 297\,K
(a) and an expanded plot from 4 to 100\,K (b) for LiV$_2$O$_4$.  The
filled circles are the neutron diffraction measurements of sample~5 by
Chmaissem~et~al.\protect\cite{Kondo1997,Chmaissem1997}  The solid curve is the
linear thermal dilation obtained from our capacitance dilatometer measurements
of sample~6, assuming $a_{\rm o}(0) = 8.22670$\,\AA.}
\label{FigAlpha2}
\end{figure}
\noindent
$\lim_{T\to 0}\alpha(T)/T = 2.00 \times 10^{-7}\,{\rm K}^{-2}$.

Shown as the solid curve in Fig.~\ref{FigAlpha2}(a) is the linear thermal
dilation expressed in terms of the lattice parameter $a_{\rm o}(T) = a_{\rm
o}(0)[1 + \int_0^T \alpha(T)\,{\rm d}T]$, where we have used our polynomial fit
to the $\alpha(T)$ data to compute $a_{\rm o}(T)$ and have set
$a_{\rm o}(0) = 8.22670$\,\AA.  The $a_{\rm o}(T)$ determined from the neutron
diffraction measurements by Chmaissem~et~al.\cite{Chmaissem1997} for a different
sample (sample~5) are  plotted as the filled circles in Fig.~\ref{FigAlpha2}. 
The two data sets are in overall agreement, and both indicate a strong decrease
in $a_{\rm o}(T)$ with decreasing $T$ below 20\,K\@.  There are differences in
detail between the two measurements at the lower temperatures as illustrated
below 100\,K in Fig.~\ref{FigAlpha2}(b), suggesting a possible sample dependence.

\begin{figure}
\epsfxsize=3.3 in
\epsfbox{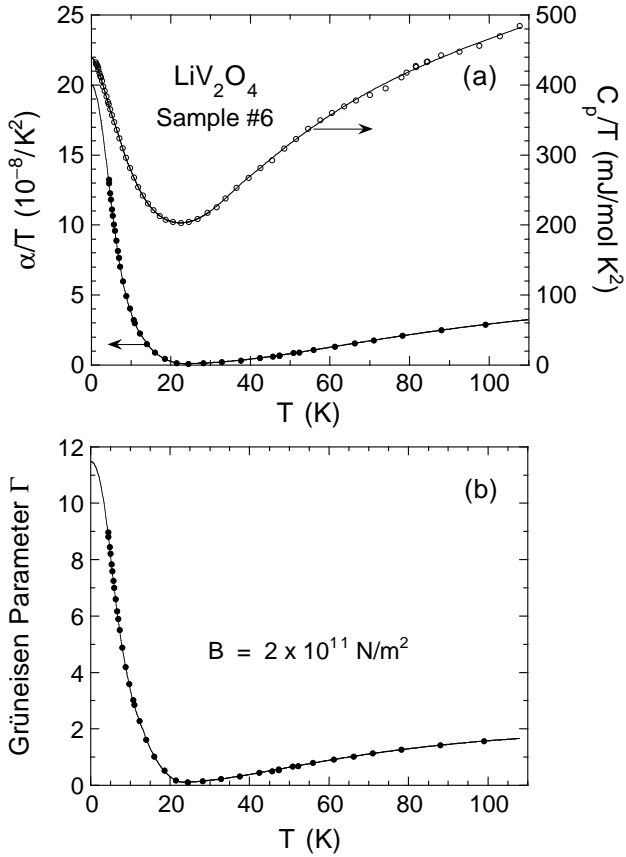}
\vglue 0.1in
\caption{(a) Comparison of the linear thermal expansion coefficient divided by
temperature $\alpha(T)/T$ (left-hand-scale) with the measured specific heat
$C_{\rm p}(T)/T$ (right-hand-scale) for LiV$_2$O$_4$ sample~6.  (b) Gr\"uneisen
parameter $\Gamma$ versus $T$, computed using Eq.~(\protect\ref{EqGrunEq}) and
the assumed value of the bulk modulus $B$ given in the figure.}
\label{FigAlpha1}
\end{figure}

Our measurement of $\alpha(T)/T$ for sample~6 is compared with the measured
$C_{\rm p}(T)/T$ for the same sample in Fig.~\ref{FigAlpha1}(a), where the
temperature dependences of these two quantities are seen to be similar.  We
infer that the strong increase in
$\alpha(T)/T$ with decreasing $T$ below $\sim 20$\,K is an electronic effect
associated with  the crossover to heavy fermion behavior.  For most materials,
\noindent
the volume thermal expansivity
$\beta = 3\alpha$ and $C_{\rm p}$ are related through the dimensionless
Gr\"uneisen parameter
$\Gamma$, with\cite{Barron1980}
\begin{equation}
\beta = \frac{\Gamma C_{\rm p}}{B_{\rm s}V_{\rm M}}~~,
\label{EqGrunEq}
\end{equation} where $B_{\rm s}$ is the adiabatic bulk modulus and $V_{\rm M}$
is the molar volume.  In this model, $\Gamma = -{\rm d\,ln}\Phi/{\rm d\,ln}V$
where $\Phi(V)$ is a characteristic energy of the system.  If independent
contributions to $C_{\rm p}$ can be identified, as assumed in
Eq.~(\ref{EqCsum:a}), a relation similar to Eq.~(\ref{EqCsum:a}) exists for the
thermal expansivity, with an independent $\Gamma$ for each contribution
\begin{equation}
\beta = \beta_{\rm e} + \beta^{\rm lat} = \frac{\Gamma_{\rm e}C_{\rm e} +
\Gamma^{\rm lat}C_{\rm p}^{\rm lat}}{B_{\rm s}V_{\rm M}}~~,
\label{EqBeta}
\end{equation} where $C_{\rm e}$ is understood to refer to measurements under
constant pressure.  For a metal,
$\Gamma_{\rm e} = {\rm d\,ln}{\cal D}^*(E_{\rm F})/{\rm d\,ln}V = {\rm
d\,ln}\gamma(0)/{\rm d\,ln}V$, and $\Gamma^{\rm lat} = -{\rm d\,ln}\theta_0/{\rm
d\,ln}V$.  Here ${\cal D}^*(E_{\rm F})$ is the mass-enhanced quasiparticle
density of states at the Fermi energy and the volume dependence of the
electron-phonon interaction is neglected.  Thus $\Gamma_{\rm e}$ is a direct
measure of the volume dependence of ${\cal D}^*(E_{\rm F})$.  For a free
electron gas, $\Gamma_{\rm e} = 2/3$.  For most real metals $\Gamma_{\rm e} =
\pm 3(2)$; {\it e.g.}, $\Gamma_{\rm e} = 0.92$ (Cu), 1.6 (Au), 1.6 (V), $-4.4$ (Sr),
$-0.2$ (Ba), 2.22 (Pd).\cite{Barron1980}

We have computed $\Gamma(T)$ for LiV$_2$O$_4$ from Eq.~(\ref{EqGrunEq}) using
the polynomial fit to our $C_{\rm p}$ data for sample~6 and using the
experimental $\alpha(T)$ data in Fig.~\ref{FigAlpha3} for this sample.  The
molar volume of LiV$_2$O$_4$ at low temperatures is given in
Table~\ref{TabLivoParams}. The bulk modulus is assumed to be $B = 200(40)$\,GPa,
which is the range found\cite{Birch1966} for the similar compounds Fe$_2$O$_3$,
Fe$_3$O$_4$, FeTiO$_3$, MgO, TiO$_2$ (rutile), the spinel prototype
MgAl$_2$O$_4$,\cite{Kruger1997} and MgTi$_2$O$_5$.\cite{Hazen1997}  The 
$\Gamma$ obtained by substituting these values into Eq.~(\ref{EqGrunEq}) is
plotted versus temperature as the filled circles in Fig.~\ref{FigAlpha1}(b). 
Interpolation and extrapolation of $\Gamma(T)$ is obtained from the polynomial 
fit to the $\alpha(T)$ data, shown by the solid curve
in Fig.~\ref{FigAlpha1}(b).  From Fig.~\ref{FigAlpha1}(b), $\Gamma\approx 1.7$
at 108\,K and decreases slowly with decreasing $T$, reaching a minimum of about
0.1 at 23\,K\@.  With further decrease in $T$, $\Gamma$ shows a dramatic
increase and we obtain an extrapolated $\Gamma(0)\approx 11.4$.  A plot of
$\Gamma$ vs.\ $T^2$ obtained from our experimental data points is linear for
$T^2 < 30$\,K$^2$, and extrapolates to 11.50 at $T = 0$, to be compared with
11.45 as calculated from the smooth fitted relations for $\alpha(T)$ and
$C_{\rm p}(T)$; this justifies the (long) extrapolation of $\alpha(T)$ to $T =
0$.  An accurate determination of the magnitude of $\Gamma$ must await the
results of bulk modulus  measurements on LiV$_2$O$_4$.  Our estimated $\Gamma(0)
\equiv \Gamma_{\rm e}(0)$ is intermediate between those of conventional
nonmagnetic metals and those of  $f$-electron heavy fermion compounds such as
UPt$_3$ ($\Gamma_{\rm e} = 71$), UBe$_{13}$ (34) and CeCu$_6$ (57) with
$\gamma(0) = 0.43,\ 0.78,$ and 1.67\,J/mol\,K$^2$,
respectively.\cite{deVisser1989}

From the expression\cite{Gopal1966} relating $C_{\rm p}$ to the specific heat at 
constant volume $C_{\rm v}$, and using our $\alpha(T)$ data and the estimate for
$B$ above,
$C_{\rm v}(T)$ of LiV$_2$O$_4$ can be considered identical with our measured
$C_{\rm p}(T)$ to within both the precision and accuracy of our measurements up
to 108\,K\@.  

\section{Theoretical Modeling: Electronic Specific Heat of
L\lowercase{i}V$_2$O$_4$}
\label{SecTheory}

\subsection{Single-Band Spin $S = 1/2$ Fermi Liquid}
\label{SecFL}

As mentioned in Sec.~\ref{SecIntro}, the high-temperature $\chi(T)$ of
LiV$_2$O$_4$ indicated a vanadium local moment with spin $S = 1/2$ and $g \sim
2$.  In the low-temperature Fermi liquid regime, for a Fermi liquid consisting
of a single parabolic band of quasiparticles with $S = 1/2$ and $N_{\rm e}$
conduction electrons per unit volume
$V$,\cite{Kittel1971,Pethick1973,Pethick1986,Baym1991} the Fermi wavevector
$k_{\rm F}$ of LiV$_2$O$_4$ assuming $N_{\rm e} = 1.5$ conduction electrons/V
atom is given in Table~\ref{TabLivoParams}.  In terms of the mass-enhanced
density of states at the Fermi energy $E_{\rm F}$ for both spin directions ${\cal
D}^*(E_{\rm F})$, the $\gamma(0)$ (neglecting electron-phonon interactions) and
$\chi(0)$ are given by
\begin{mathletters}
\label{EqDOS:all}
\begin{equation}
\gamma(0) = \frac{\pi^2k_{\rm B}^2}{3} {\cal D}^*(E_{\rm F})~~,\label{EqDOS:a}
\end{equation}
\begin{equation}
\chi(0) = \frac{g^2\mu_{\rm B}^2}{4} \frac{{\cal D}^*(E_{\rm F})}{1 + F_0^{\rm
a}}~~,
\label{EqDOS:b}
\end{equation}
\end{mathletters}
where $F_0^{\rm a}$ is a Landau Fermi liquid parameter and $1/(1 + F_0^{\rm a}) =
1 - A_0^{\rm a}$ is the Stoner enhancement factor.  The Fermi liquid scattering
amplitudes $A_\ell^{\rm a,s}$ are  related to the Landau parameters $F_\ell^{\rm
a,s}$ by $A_\ell^{\rm a,s} = F_\ell^{\rm a,s}/[1 + F_\ell^{\rm a,s}/(2\ell +
1)]$.  The superscripts ``a'' and ``s'' refer to spin-asymmetric and
spin-symmetric interactions, respectively.  Using Eq.~(\ref{EqDOS:a}) and  the
$k_{\rm F}$ value in Table~\ref{TabLivoParams}, the experimental value of
$\gamma(0)$ for LiV$_2$O$_4$ in Eq.~(\ref{EqGam(0):a}) yields the effective mass
$m^*$, Fermi velocity $v_{\rm F}$, $E_{\rm F}$, Fermi temperature $T_{\rm F}$
and ${\cal D}^*(E_{\rm F})$ for LiV$_2$O$_4$ given in
Table~\ref{TabLivoParams}.  From Eqs.~(\ref{EqRW}) and~(\ref{EqDOS:all}), the
Wilson ratio\cite{Wilson1975} $R_{\rm W}$ is expressed as
\begin{equation}
R_{\rm W} = \frac{1}{1 + F_0^{\rm a}} = 1 - A_0^{\rm a}~~.
\label{EqWilRat}
\end{equation}
Substituting the experimental $\chi(0.4$--2\,K) = 0.0100(2) cm$^3$/mol
(Ref.~\onlinecite{Kondo1997}) and $\gamma(0)$ in Eq.~(\ref{EqGam(0):a}) for
LiV$_2$O$_4$ into Eq.~(\ref{EqRW}) assuming $g = 2$ yields
\begin{equation}
R_{\rm W} = 1.71(4)~~.
\label{EqRexp}
\end{equation}
This $R_{\rm W}$ value is in the range of those found for many conventional as
well as $f$-electron HF and IV compounds.\cite{HFIV}  The $R_{\rm W}$ value in
Eq.~(\ref{EqRexp}) yields from Eq.~(\ref{EqWilRat})
\begin{equation}
F_0^{\rm a} = -0.42,~~~~A_0^{\rm a} = -0.71~~.
\label{EqLanParams2}
\end{equation}
In Fermi liquid theory, a temperature dependence is often computed for $C_{\rm
e}$ at low temperatures having the form\cite{Pethick1973,Pethick1986,Baym1991} 

\begin{table}
\caption{Parameters for LiV$_2$O$_4$.  Abbreviations: formula weight FW; lattice
parameter 
$a_0$;\protect\cite{Kondo1997,Chmaissem1997} (formula units/unit cell) $Z$;
theoretical mass density $\rho^{\rm calc}$; molar volume $V_{\rm M}$, itinerant
electron concentration $N_{\rm e}/V$; Fermi wavevector
$k_{\rm F}  = (3\pi^2N_{\rm e}/V)^{1/3}$; effective mass $m^*$ (free electron
mass $m_{\rm e}$); Fermi velocity $v_{\rm F} = \hbar k_{\rm F}/m^*$; Fermi
energy $E_{\rm F} = \hbar^2 k_{\rm F}^2/2 m^*$; Fermi temperature $T_{\rm F} =
E_{\rm F}/k_{\rm B}$; mass-enhanced density of states at $E_{\rm F}$ for both
spin directions, ${\cal D}^*(E_{\rm F}) = 3 N_{\rm e}/ (2 E_{\rm F}) = m^*
k_{\rm F} V/(\pi^2\hbar^2)$.}
\begin{tabular}{ld} Property & value \\
\hline FW & 172.82 g/mol \\
$a_0$(12\,K) & 8.2269\,\AA \\
$Z$ & 8 \\
$\rho^{\rm calc}(12\,{\rm K})$ & 4.123\,g/cm$^3$ \\
$V_{\rm M}$ &  41.92\,cm$^3$/mol \\
$N_{\rm e}/V$ & 4.310\,$\times 10^{22}\,{\rm cm^{-3}}$ \\
$k_{\rm F}$ & 1.0847\,\AA$^{-1}$ \\
$m^*/m_{\rm e}$ & 180.5 \\
$v_{\rm F}$ & 6.96\,$\times 10^5$\,cm/s \\
$E_{\rm F}$ & 24.83\,meV \\
$T_{\rm F}$ & 288.1\,K \\
${\cal D}^*(E_{\rm F})$ & 90.6\,states/eV(V\,atom) \\
\end{tabular}
\label{TabLivoParams}
\end{table}
\begin{equation}
C_{\rm e}(T) = \gamma(0)T + \delta T^3 \ln\Big(\frac{T}{T_0}\Big) + {\cal
O}(T^3)~~,
\label{EqCeFL}
\end{equation}
where $\gamma(0)$ is given by Eq.~(\ref{EqDOS:a}) and $T_0$ is a
scaling or cutoff temperature.  Engelbrecht and Bedell\cite{Engelbrecht1995}
considered a model of a single-band Fermi liquid with the microscopic constraint
of a local (momentum-independent) self-energy, where the interactions are
mediated by the quasiparticles themselves (in the small momentum-transfer
limit).  They find that only $s$-wave ($\ell = 0$) Fermi-liquid parameters can
be nonzero and that the $\delta$ coefficient in Eq.~(\ref{EqCeFL}) is
\begin{equation}
\delta_{\rm EB} = \frac{3\pi^2}{5}\frac{\gamma(0)}{T_{\rm F}^2}\bigl(A_0^{\rm
a}\bigr)^2\biggl(1 -
\frac{\pi^2}{24}A_0^{\rm a}\biggr)~~,
\label{EqDeltaEB}
\end{equation} where $|A_0^{\rm a,s}| \leq 1$ and $-\frac{1}{2} \leq F_0^{\rm
a,s} < \infty$.  Within their model, neither ferromagnetism  nor
phase-separation can occur.  For $F_0^{\rm a} < 0$, the only potential
instability is towards antiferromagnetism and/or a metal-insulator transition;
in this case they find $1 \leq R_{\rm W} \leq 2$.  For $F_0^{\rm a} > 0$, a BCS
superconducting  state is possible and $R_{\rm W} < 1$.  The value of $F_0^{\rm
a}$ for  LiV$_2$O$_4$ in Eq.~(\ref{EqLanParams2}) is within the former range of
this theory.  

Auerbach and Levin\cite{Auerbach1986} and Millis {\it et
al.}\cite{Millis1987,Millis1987b} formulated a Fermi-liquid theory of heavy
electron compounds at low temperatures on the basis of a microscopic Kondo
lattice model.  The large enhancement of $m^*$ arises from the spin entropy of
the electrons on the magnetic-ion sites ({\it i.e.}, spin
fluctuations).\cite{Millis1987}  The Wilson ratio is $R_{\rm W} \sim 1.5$ and a
$T^3 \ln T$ contribution to $C_{\rm e}(T)$ is found.  The origin of this latter
term is not ferromagnetic spin fluctuations
(``paramagnons''),\cite{Auerbach1986} but is rather electron density
fluctuations and the screened long-range Coulomb interaction.\cite{Millis1987} 
The coefficient
$\delta_{\rm M}$ of the $T^3 \ln T$ term found by  Millis\cite{Millis1987} is
$\delta_{\rm M} = \pi^2k_{\rm B}^4V(1 - \pi^2/12)/5(\hbar v_{\rm F})^3$, which
may be rewritten  as
\begin{equation}
\delta_{\rm M} = \frac{3\pi^2\gamma(0)}{20T_{\rm F}^2}\Big(1 -
\frac{\pi^2}{12}\Big)~~.
\label{EqDeltaMillis}
\end{equation}

Using the values $\gamma(0) = 427$\,mJ/mol\,K$^2$ [Eq.~(\ref{EqGam(0):a})],
$T_{\rm F}$ = 288\,K (Table~\ref{TabLivoParams}) and $A_0^{\rm a}$ in
Eq.~(\ref{EqLanParams2}), Eqs.~(\ref{EqDeltaEB})  and~(\ref{EqDeltaMillis})
respectively predict
\begin{equation}
\delta_{\rm EB} = 0.0199\,\frac{\rm mJ}{\rm mol\,K^4}~,~~~~\delta_{\rm M} =
0.00135\,\frac{\rm mJ}{\rm mol\,K^4}~~.
\label{EqDelCalcs}
\end{equation}

We have fitted our low-temperature $C_{\rm e}(T)/T$ data for LiV$_2$O$_4$
sample~3 by the expression
\begin{equation}
\gamma(T) \equiv \frac{C_{\rm e}(T)}{T} =  \gamma(0) +
\delta\,T^2\ln \biggl(\frac{T}{T_0}\biggr) + \varepsilon T^3~~,
\label{EqFLCT2}
\end{equation} initially with $\varepsilon = 0$.  The fit parameters
$\gamma(0),\ \delta$ and $T_0$ were found to depend on the fitting temperature
range above 1\,K chosen, and are sensitive to the precision of the data.  The
parameters obtained for 1--3\,K and 1--5\,K fits were nearly the same, but
changed when the upper 
\begin{figure}
\epsfxsize=3.2 in
\epsfbox{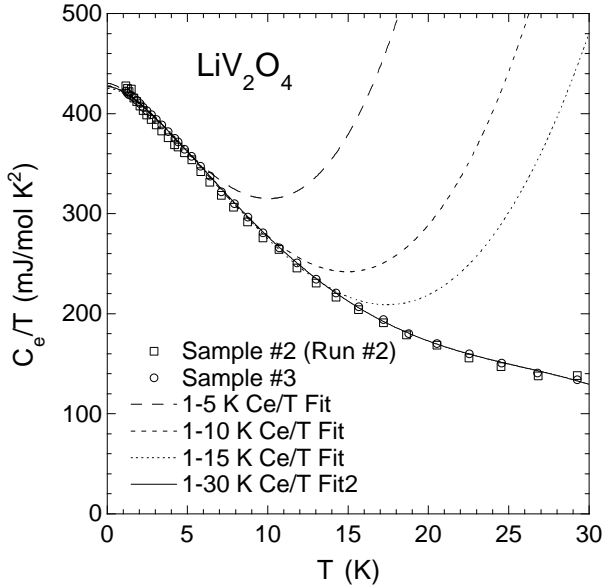}
\vglue 0.1in
\caption{Electronic specific heat $C_{\rm e}$ divided by temperature $T$ for
LiV$_2$O$_4$ samples~2 (run~2) and~3 {\it vs.}\ $T$.  The dashed curves are fits
to the 1--5\,K, 1--10\,K and 1--15\,K data for sample~3 by the spin-fluctuation
Fermi liquid model, Eq.~(\protect\ref{EqFLCT2}) with $\varepsilon = 0$, whereas
the solid curve is a 1--30\,K fit assuming $\varepsilon \neq 0$.\label{FigCTFit}}
\end{figure}
\noindent
limit to the fitting range was increased to 10 and
15\,K\@.  The fits for the 1--5\,K, 1--10\,K and 1--15\,K fitting ranges are
shown in Fig.~\ref{FigCTFit}, along with the $C_{\rm e}(T)/T$ data for sample~2
(run~2).  As a check on the fitting parameters, we have also fitted the $C_{\rm
e}(T)/T$ data for sample~3 by Eq.~(\ref{EqFLCT2}) with $\varepsilon$ as an
additional fitting parameter.  The fit for the 1--30\,K range is plotted as the
solid curve in Fig.~\ref{FigCTFit}.  Since the fits for the smaller $T$ ranges
with  $\varepsilon = 0$ and for the larger ranges with $\varepsilon \neq 0$
should give the most reliable parameters, we infer from the fit parameters
for all ranges that the most likely values of the parameters and their error bars
are
\begin{equation}
\gamma(0) = 428(2)\,{\rm \frac{mJ}{mol\,K^2}},~~~\delta = 1.9(3) {\rm
\frac{mJ}{mol\,K^4}}~~.
\label{EqCTParams}
\end{equation}
The parameters in Eq.~(\ref{EqCTParams}) are very similar to those obtained using
the same type of fit to $C_{\rm p}(T)/T$ data for the heavy fermion
superconductor UPt$_3$ with $T_{\rm c} = 0.54$\,K,\cite{Stewart1984} for which
$\gamma(0) = 429$--450\,mJ/mol\,K$^2$ and $\delta =
1.99$\,mJ/mol\,K$^4$.\cite{Stewart1984,Trinkl1996}   Our $T_{\rm F}$ and
$m^*/m_{\rm e}$ values for LiV$_2$O$_4$ (288\,K and 181,
Table~\ref{TabLivoParams}) are also respectively very similar to those of
UPt$_3$ (289\,K and 178).\cite{Pethick1986}

The experimental $\delta$ value in Eq.~(\ref{EqCTParams}) is a factor of $\sim
10^2$ larger than $\delta_{\rm EB}$ and $\sim 10^3$ larger than $\delta_{\rm M}$
predicted in Eq.~(\ref{EqDelCalcs}).  A similar large [${\cal O}(10^2$--$10^3)$]
discrepancy was found by Millis for the $\delta$ coefficient for
UPt$_3$.\cite{Millis1987}  As explained by Millis,\cite{Millis1987} the large
discrepancy between his theory and experiment may arise because the calculations
are for a single parabolic band, an assumption which may not be applicable to
the real materials.  However, he viewed the most likely reason to be that his
calculations omit some effect important to the thermodynamics such as
antiferromagnetic spin fluctuations.\cite{Millis1987}  In this context, it is
possible that the magnitude of $\delta$ predicted by one of the above two
theories is correct, but that terms higher order in $T$ not calculated by the
theory are present which mask the $T^3 \ln T$ contribution over the temperature
ranges of the fits;\cite{Ishigaki1996a} in this case the large experimental
$\delta$ value would be an artifact of force-fitting the data by
Eq.~(\ref{EqFLCT2}).  Indeed, we found that the fits were unstable, {\it i.e.},
depended on the temperature range fitted ({\it cf.}\ Fig.~\ref{FigCTFit}).  In
addition, the applicability of the theory of Millis\cite{Millis1987} to
LiV$_2$O$_4$ is cast into doubt by the prediction that the Knight shift at a
nucleus of an atom within the conduction electron sea (not a ``magnetic'' atom)
``would be of the same order of magnitude as in a normal metal, and would not
show the mass enhancement found in
$\chi$.''\cite{Millis1987b}  In fact, the Knight shift of the $^7$Li nucleus in
LiV$_2$O$_4$ for $T \sim 1.5$--10\,K is about
0.14\%,\cite{Kondo1997,Mahajan1998,Onoda1997,Fujiwara1997} which is about 6000
times  larger than the magnitude (0.00024\%) found\cite{Dalton1994} at room
temperature for the
$^7$Li Knight shift in LiTi$_2$O$_4$.  Similarly, the $^7$Li 1/$T_1 T$ from 1.5
to 4\,K in the highest-purity LiV$_2$O$_4$ samples is about
2.25\,s$^{-1}$K$^{-1}$,\cite{Kondo1997,Mahajan1998} which is about 6000 times
larger than the value of $3.7 \times 10^{-4}$\,s$^{-1}$K$^{-1}$
found\cite{Dalton1994} at 160\,K in LiTi$_2$O$_4$, where $T_1$ is the $^7$Li
nuclear spin-lattice relaxation time.

\subsection{Quantum-Disordered Antiferromagnetically Coupled Metal}
\label{SecMillis}

  The antiferromagnetic (AF) Weiss temperature of LiV$_2$O$_4$ from $\chi(T)$
measurements is
$|\theta| = 30$--60\,K, yet the pure system exhibits neither static
antiferromagnetic AF nor spin-glass order above
0.02\,K.\cite{Kondo1997,Kondo1998}  A  possible explanation is that the ground
state is disordered due to quantum fluctuations.  We consider here the
predictions for
$C_{\rm e}(T)$ of one such theory.  A universal contribution to the temperature
dependence of
$C_{\rm e}$ of a three-dimensional (3D) metal with a control parameter $r$ near
that required for a zero-temperature AF to quantum-disordered phase transition,
corresponding to dynamical exponent
$z = 2$, was calculated by Z\"ulicke and Millis,\cite{Zulicke1995} which
modifies the Fermi liquid prediction in Eq.~(\ref{EqCeFL}).  Upon increasing $T$
from $T = 0$ in the quantum-disordered region, the system crosses over from the
quantum disordered to a classical regime.  The same scaling theory predicts that
the low-$T$ spin susceptibility is given by
$\chi(T) = \chi(0) + A T^{3/2}$, where the constant $A$ is not determined by the
theory.\cite{Ioffe1995}

Z\"ulicke and Millis found the electronic specific heat to be given
by\cite{Zulicke1995}
\begin{mathletters}
\label{EqGamCalcMillis:all}
\begin{equation}
\frac{C_{\rm e}}{T} = \gamma_0 - \frac{\alpha  R
N_0\sqrt{r}}{6T^*}\,F\Big(\frac{T}{rT^*}\Big)~~,
\label{EqGamCalcMillis:a}
\end{equation}
\begin{equation} F(x) = \frac{3\sqrt{2}}{\pi^2}\int_0^\infty {
dy\,\frac{y^2}{\sinh^2y}\sqrt{1 + \sqrt{1 + 4 x^2 y^2}} }~~.
\label{EqGamCalcMillis:b}
\end{equation}
\end{mathletters} Here, $\gamma_0$ is the (nonuniversal) electronic specific
heat coefficent at $T = 0$ in the usual Fermi liquid theory [$\gamma(0)$ above],
$T^*$ is a characteristic temperature and $N_0$ is the number of components of
the bosonic order parameter which represents the ordering field: $N_0$ = 3, 2, 1
for Heisenberg, {\em XY} and Ising symmetries, respectively.  The number
$\alpha$ is not determined by the scaling theory but is expected to be on the
order of the number of conduction electrons per formula unit; thus for
LiV$_2$O$_4$, we expect $\alpha \sim 3$.  We have defined
$F(x)$ such that $F(0) = 1$.  The variable $r$ is expected to be temperature
dependent, but this temperature dependence cannot be evaluated without
ascertaining the value of an additional parameter $u$ in the theory from, {\it
e.g.}, measurements of the pressure dependence of $C_{\rm e}(T)$; here, we will
assume $r$ to be a constant.\cite{Assump}  From Eq.~(\ref{EqGamCalcMillis:a}),
the $T=0$ value of
$\gamma$ in the absence of quantum fluctuations is reduced by these
fluctuations, and  the measured $\gamma(0)$ is
\begin{equation}
\gamma(0) = \gamma_0  - \frac{\alpha R N_0\sqrt{r}}{6T^*}~~.
\label{EqGamMeas}
\end{equation}

\begin{figure}
\epsfxsize=3.3 in
\epsfbox{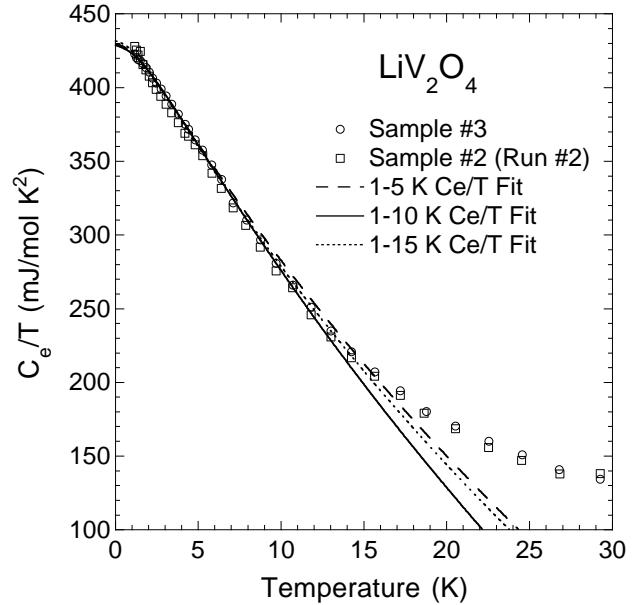}
\vglue 0.1in
\caption{Electronic specific heat divided by temperature $C_{\rm e}/T$ {\it
vs.}\ $T$ for LiV$_2$O$_4$ samples~2 (run~2) and~3.  Fits to the data for
sample~3  by the theory of Z\"ulicke and Millis,\protect\cite{Zulicke1995}
Eqs.~(\protect\ref{EqGamCalcMillis:all}), are shown for the fitting ranges
1--5\,K (long-dashed curve), 1--10\,K (solid curve) and 1--15\,K (short-dashed
curve).\label{FigMillisFit}}
\end{figure}

We fitted our $C_{\rm e}/T$ {\em vs.}\ $T$ data for LiV$_2$O$_4$ sample~3  by
Eqs.~(\ref{EqGamCalcMillis:all}), assuming $N_0$ =~3.  The fitting parameters 
were $\gamma_0,\ \alpha,\ r$ and~$T^*$; the $\gamma(0)$ value is then obtained 
from Eq.~(\ref{EqGamMeas}).  The 1--20\,K and larger ranges did not give
acceptable fits.  The fits for the 1--5, 1--10 and 1--15\,K fitting ranges are
shown in Fig.~\ref{FigMillisFit}.  From these fits, we infer the parameters and
errors
\begin{eqnarray}
\gamma_0 & = & 800(50)\,\frac{\rm mJ}{\rm mol\,K^2}~,~~~\alpha = 2.65(9)~,~~~r =
0.40(6)~,\nonumber\\ T^* & = & 18.9(4)\,{\rm K}~,~~~\gamma(0) = 430(1)\,{\rm
mJ/mol\,K^2}~.
\label{EqMillisParams}
\end{eqnarray}
Within the context of this theory, quantum fluctuations reduce the observed
$\gamma(0)$ by about a factor of two compared with the value $\gamma_0$ in the
absence of these fluctuations.  The value of $\alpha$ is close to the nominally
expected value $\sim 3$ mentioned above.  The relatively large value of $r$
indicates that LiV$_2$O$_4$ is not very close to the quantum-critical point, and
therefore predicts that long-range AF order will not be induced by small changes
in external conditions (pressure) or composition.  The former prediction cannot
be checked yet because the required experiments under pressure have not yet been
done.  The latter expectation is consistent with the data available so far. 
Magnetic defect concentrations on the order of 1\% do induce static magnetic
ordering below $\sim 0.8$\,K, but this ordering is evidently of the short-range
spin-glass type.\cite{Kondo1997}  Substitution of Zn for Li in
Li$_{1-x}$Zn$_x$V$_2$O$_4$ induces spin-glass ordering for $0.2 \lesssim x
\lesssim 0.9$ but long-range AF ordering does not occur until $0.9
\lesssim x \leq 1.0$.\cite{Ueda1997}  Finally, two caveats regarding the fits
and discussion in this section are in order.  The first is that (unknown)
corrections of order $(T/T^*)^2$ and $r^1$ to the theory of Z\"ulicke and
Millis\cite{Zulicke1995} exist but have not been included in the prediction in
Eqs.~(\ref{EqGamCalcMillis:all}); incorporating these corrections may alter the
parameters obtained from fits to experimental data.\cite{Millis1997}  The second
caveat is that the theory 
\noindent
may need modification for compounds such as
LiV$_2$O$_4$ in which geometric frustration for AF ordering exists in the
structure.\cite{Millis1997}

\subsection{Spin-1/2 Kondo Model}
\label{SecS1/2Kondo}

Calculations of the impurity spin susceptibility $\chi(T)$ and/or impurity
electronic contribution $C_{\rm e}(T)$ to the specific heat for the $S = 1/2$
Kondo model were carried out by Wilson\cite{Wilson1975} and
others\cite{Krishna1980a,Krishna1980b,Oliveira1981,Rajan1982,Rajan1983,Desgranges1982,Tsvelick1983,Jerez1997}
using different techniques.  Both $\chi(T)$ and $C_{\rm e}(T)$ depend only on the
scaling parameter $T/T_{\rm K}$, where $T_{\rm K}$ is the Kondo temperature
(here, we use Wilson's definition\cite{Wilson1975}).  The impurity $\chi(T)$ is
predicted to be Curie-Weiss-like at temperatures high compared with $T_{\rm K}$,
and to level out at a constant high value for $T \lesssim 0.1\,T_{\rm K}$ due to
the formation of a singlet ground state.

In the limit of zero temperature, one has\cite{Rajan1983}
\begin{equation}
\gamma(T = 0) = \frac{\pi W N k_{\rm B}}{6 T_{\rm K}}~~,
\label{EqGam00}
\end{equation} where $N$ is the number of impurity spins.  The Wilson
number\cite{Wilson1975} $W$ is given by\cite{Andrei1981,Rasul1984}
\begin{equation} W = \gamma{\rm e}^{1/4}\pi^{-1/2} \approx 1.290\,268\,998~~,
\label{EqW}
\end{equation} where $\ln\gamma \approx 0.577\,215\,664\,902$ is Euler's
constant.  Setting $N =  N_{\rm A}$, Avogadro's number, one obtains from
Eqs.~(\ref{EqGam00}) and~(\ref{EqW}) the electronic specific heat coefficient
per mole of impurities
\begin{equation}
\gamma(0) = \frac{\pi WR }{6T_{\rm K}} =
\frac{5.61714\,{\rm J/mol\,K}}{T_{\rm K}}~~.
\label{EqGam0Calc}
\end{equation}

To characterize the $T$ dependence of $C_{\rm e}$, we utilized accurate
numerical calculations using the Bethe ansatz by Jerez and
Andrei.\cite{Jerez1997} The calculated
$C_{\rm e}(T)$ shows a maximum, max[$C_{\rm e}(T)/Nk_{\rm B}] = 0.177275$, which
occurs at
$T^{\rm max}/T_{\rm K} = 0.6928$.  The  calculations were fitted by the
expressions
\begin{mathletters}
\label{EqC(T):all}
\begin{equation}
\frac{C_{\rm e}(T)}{N k_{\rm B}} = f(t)~~,\label{EqC(T):a}
\end{equation}
\begin{equation}
\frac{C_{\rm e}(T)}{N k_{\rm B} T/T_{\rm K}} = g(t) \equiv
\frac{f(t)}{t}~~,\label{EqC(T):b}
\end{equation}
\begin{equation} f(t) = \bigg(\frac{\pi W}{6}\bigg) \frac{t(1 + a_1 t + a_2 t^2
+ a_3 t^3 + a_4 t^4)}{1 + a_5 t + a_6 t^2 + a_7 t^3 + a_8 t^4 + a_9
t^5}~~,\label{EqC(T):c}
\end{equation}
\end{mathletters} where $t \equiv T/T_{\rm K}$ and the coefficients $a_n$ for
the two types of fits are given in Table~\ref{TableVI} for the fitting range
$0.001 \leq t \leq 100$.   Equations (\ref{EqC(T):all}) incorporate the
zero-temperature limit in Eqs.~(\ref{EqGam00}--\ref{EqGam0Calc}). The maximum
(rms) deviations of the $C_{\rm e}(T)$ fit from the calculated numerical data
are 0.011\% (0.0035\%) for $0
\leq t \leq 3$ and 0.031\% (0.021\%) for $3 \leq t
\leq 10$ but then progressively deteriorate to 0.48\% (0.14\%) in the region $10
\leq t
\leq 92$.  The corresponding deviations for the $C_{\rm e}(T)/T$ fit are 0.0044\%
(0.00091\%), 0.031\% (0.017\%) and 5.1\% (1.6\%). 

\begin{figure}
\epsfxsize=3.2 in
\epsfbox{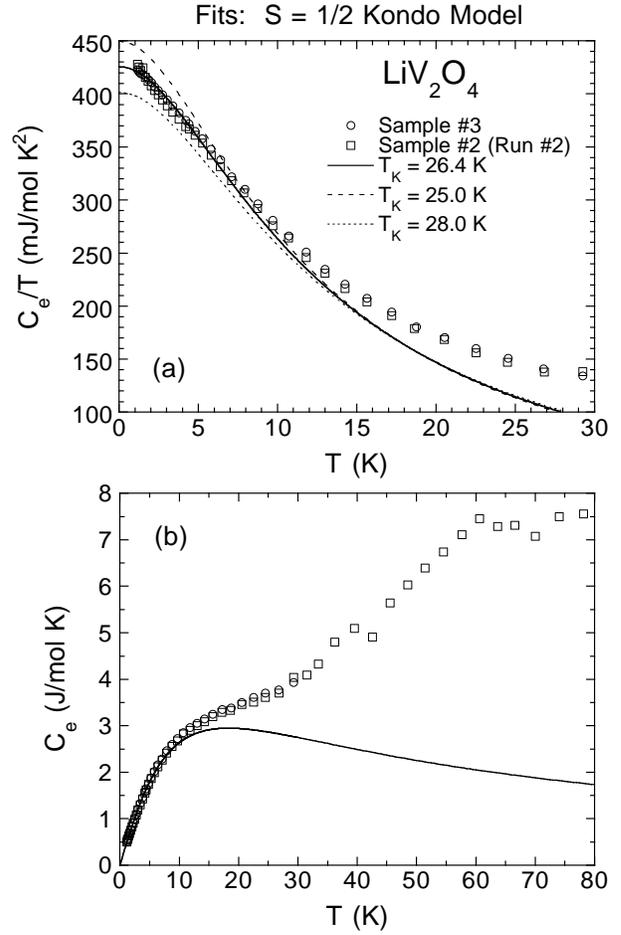}
\vglue 0.1in
\caption{(a) Electronic specific heat divided by temperature $C_{\rm e}(T)/T$
data for LiV$_2$O$_4$ samples~2 (run~2) and~3 below 30\,K (open symbols) and a
fit (solid curve) of the data for sample~3 by the $S = 1/2$ Kondo model,
Eqs.~(\protect\ref{EqC(T):b}) and~(\protect\ref{EqC(T):c}), for a Kondo
temperature $T_{\rm K}$ = 26.4\,K\@.  Shown for comparison are the predictions
for $T_{\rm K}$ = 25.0\,K (long-dashed curve) and 28.0\,K (short-dashed curve). 
(b) The same data and the fit with $T_{\rm K}$ = 26.4\,K in a plot of $C_{\rm
e}$ {\it vs.}\ $T$ up to 80\,K\@.\label{FigCpFit1}}
\end{figure}

The experimental $C_{\rm e}(T)/T$ data for LiV$_2$O$_4$ sample~3 were
least-square fitted from 1.2 to 5\,K by Eqs.~(\ref{EqC(T):b})
and~(\ref{EqC(T):c}),\cite{DigitizedC} yielding $T_{\rm K}$, and then
$\gamma(0)$ from Eq.~(\ref{EqGam0Calc}):
\begin{equation}
T_{\rm K} = 26.4(1)\,{\rm K}~,~~~\gamma(0) =~426(2)\,{\rm mJ/mol\,K^2}~.
\end{equation}
The fit is shown in Fig.~\ref{FigCpFit1} as the solid curves.  For comparison,
also shown in Fig.~\ref{FigCpFit1}(a) are the predictions for $T_{\rm K}$~= 25\,K
and 28\,K\@.  Unfortunately, despite the good agreement of the theory for
$T_{\rm K} = 26.4$\,K with our measured $C_{\rm e}(T)$ at low $T$, the $S = 1/2$
Kondo model prediction for $\chi(T)$ qualitatively disagrees with the observed
temperature dependence at low $T$.\cite{Kondo1998}  This difficulty of
self-consistently fitting the $C_{\rm e}(T)$ and $\chi(T)$ data is a problem 
\begin{table}
\caption{Coefficients $a_n$ in Eq.~(\protect\ref{EqC(T):c}) in the fits to the
theoretical prediction for the specific heat vs.\ temperature of the $S = 1/2$
Kondo model by Jerez and Andrei.\protect\cite{Jerez1997}}
\begin{tabular}{cdd}
$a_n$ & $C(T)$ Fit & $C(T)/T$ Fit\\
\hline
$a_1$ & 9.1103933 & 6.8135534\\
$a_2$ & 30.541094 & 21.718636\\
$a_3$ & 2.1041608 & 2.3491812\\
$a_4$ & 0.0090613513 & 0.017533911\\
$a_5$ & 9.1164094 & 6.8158433\\
$a_6$ & 36.143206 & 27.663307\\
$a_7$ & 67.91795 & 48.229552\\
$a_8$ & 53.509135 & 40.216156\\
$a_9$ & 1.7964377 & 2.4863342\\
\end{tabular}
\label{TableVI}
\end{table}
\noindent
we have encountered in all our attempts so far to fit our data for both
measurements over any extended temperature range by existing theory (see also
the next section).
\subsection{Local Moment High-Temperature Description}
\label{SecHTS}

As discussed above, the $\chi(T)$ data for LiV$_2$O$_4$ suggest that at
high temperatures a local moment description in which the moments are
antiferromagnetically coupled with Weiss temperature $\theta \sim -30$ to
$-60$\,K may be applicable.\cite{Kondo1997,Kondo1998}  Accordingly, we have
calculated the magnetic specific heat $C_{\rm m}(T)$ for localized moments on
the octahedral ({\it B}) sublattice of the {\it A}[{\it B}$_2$]O$_4$ spinel
structure assuming nearest-neighbor AF Heisenberg interactions using the general
high-temperature series expansion (HTSE) results of Rushbrooke and
Wood.\cite{Rushbrooke1958}  The Hamiltonian is ${\cal H} = J
\sum_{<ij>} \bbox{S}_i\cdot \bbox{S}_j$, where the sum is over all exchange bonds
and the exchange constant $J > 0$  corresponds to AF interactions.  In terms of
this Hamiltonian, $\theta = -zJS(S + 1)/3$, where $z = 6$ is the coordination
number for the {\it B} sublattice of the spinel structure.  The above range of
$\theta$ then gives $J/k_{\rm B} = 20$--40\,K assuming $S = 1/2$.  The
general HTSE  prediction is\cite{Rushbrooke1958}
\begin{equation}
\frac{C_{\rm m}(T)}{Nk_{\rm B}} = \frac{z[S(S + 1)]^2}{6t^2}\bigg[1 + \sum_{n =
1}^{n^{\rm max}}{\frac{c_n(S)}{t^n}\bigg]}~~,
\label{EqHTS}
\end{equation}
where $t \equiv k_{\rm B}T/J$ and the coefficients $c_n$ depend in general on the
spin-lattice structure in addition to $S$.  The coefficients $c_n$ for the {\it B}
sublattice of the spinel structure with $S = 1/2$ and $S = 1$ up to the  maximum
available $n^{\rm max} = 5$ are given in Table~\ref{TableVII}.  The  predictions
for $C_{\rm m}$ versus scaled-temperature $k_{\rm B}T/[JS(S + 1)]$ with 
$n^{\rm
max}$ = 5 are very similar for $S = 1/2$ and $S = 1$.  A comparison 

\begin{figure}
\epsfxsize=3.1 in
\epsfbox{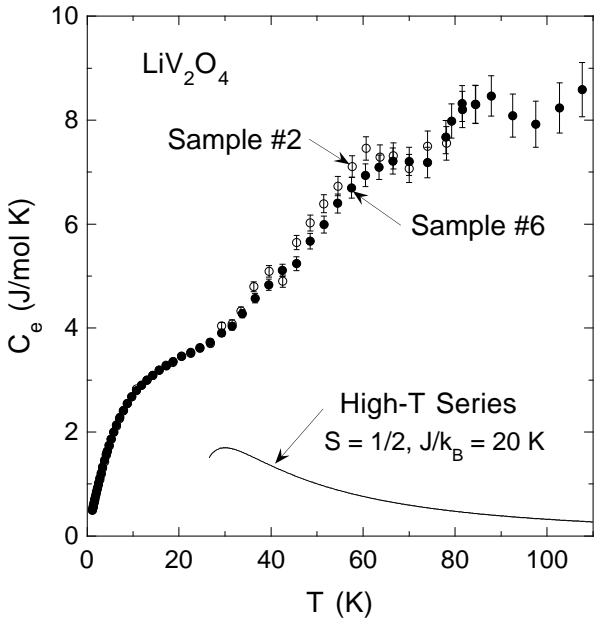}
\vglue 0.1in
\caption{Comparison of the high temperature series expansion prediction for the
magnetic specific heat $C_{\rm m}(T)$ of the {\it B} sublattice of the {\it
A}[{\it B}$_2$]O$_4$ spinel structure assuming $S = 1/2$, $J/k_{\rm B} = 20$\,K
and $n^{\rm max} = 5$, given by Eq.~(\protect\ref{EqHTS}) with $c_n$
coefficients in Table~\protect\ref{TableVII}, with the experimental $C_{\rm
e}(T)$ data for LiV$_2$O$_4$ sample~2 (run~2) and sample~6 from
Fig.~\protect\ref{FigCe}(a).\label{FigHTSFit}}
\end{figure}

\begin{table}
\caption{Coefficients $c_n$ in Eq.~(\protect\ref{EqHTS}) for the high
temperature series expansion for the magnetic specific heat of the {\it B}
sublattice of the spinel structure, for the indicated values of spin $S$.}
\begin{tabular}{ccc}
$n$ & $S = 1/2$ & $S = 1$\\
\hline 1  &  $-1/2$  &  $-$13/6\\
 2  &  $-23/16$  &  $-$3\\
 3  &  $65/48$  &  715/36\\
 4  &  $1183/768$  &  $-$4421/324\\
 5  & $-18971/7680$  &  $-$670741/6480\\
\end{tabular}
\label{TableVII}
\end{table}
\noindent
of the $C_{\rm
m}(T)$ predictions for $n^{\rm max}$ = 0 to 5 indicates that the calculations for
$n^{\rm max} = 5$ are accurate for $k_{\rm B}T/[JS(S+1)] \gtrsim 2.5$, a $T$ range
with a lower limit slightly above the temperatures at which broad maxima occur.

In Fig.~\ref{FigHTSFit} the HTSE prediction of $C_{\rm m}(T)$ for the {\it B}
sublattice of the spinel structure with $n^{\rm max} = 5$, $S = 1/2$ and
$J/k_{\rm B} = 20$\,K in Eq.~(\ref{EqHTS}) is compared with the experimental
$C_{\rm e}(T)$ data for LiV$_2$O$_4$ samples~2 and~6 from Fig.~\ref{FigCe}(a). 
The HTSE $C_{\rm m}(T)$ has a much lower magnitude than the data and a
qualitatively different temperature dependence.  From Eq.~(\ref{EqHTS}), changing
$J$ just scales the curve with $T$.  Thus the local moment picture is in severe
disagreement with our $C_{\rm e}(T)$ measurements, despite the excellent agreement
between the corresponding HTSE
$\chi(T)$ prediction and the $\chi(T)$ data from 50--100\,K to
400\,K\@.\cite{Kondo1997,Kondo1998}  

\section{Summary and Concluding Remarks}
\label{SecConcl}

We have presented $C_{\rm p}(T)$ data for LiV$_2$O$_4$ sample~6 which extend our
previous measurements\cite{Kondo1997} up to 108\,K\@.  We have also presented
$C_{\rm p}(T)$ data for the isostructural superconducting compound LiTi$_2$O$_4$
($T_{\rm c} = 11.8$\,K) up to 108\,K which complement our earlier
data\cite{Kondo1997} on the isostructural nonmagnetic insulator
Li$_{4/3}$Ti$_{5/3}$O$_4$.  We concluded here that the lattice contribution
$C^{\rm lat}(T)$ to $C_{\rm p}(T)$ for LiTi$_2$O$_4$ provides the most reliable
estimate of the $C^{\rm lat}(T)$ for LiV$_2$O$_4$, and we then extracted the
electronic contribution
$C_{\rm e}(T)$ to $C_{\rm p}(T)$ of LiV$_2$O$_4$ from 1.2 to 108\,K\@. 
Inelastic neutron  scattering measurements of the lattice dynamics and spin
excitations would be very useful in interpreting the measurements presented
here.  It will be important to determine whether or not there exist significant
differences in the lattice dynamics of LiV$_2$O$_4$ and LiTi$_2$O$_4$; in our
data analyses and modeling, we have assumed that these compounds are similar in
this respect.  

\begin{figure}
\epsfxsize=3.3 in
\epsfbox{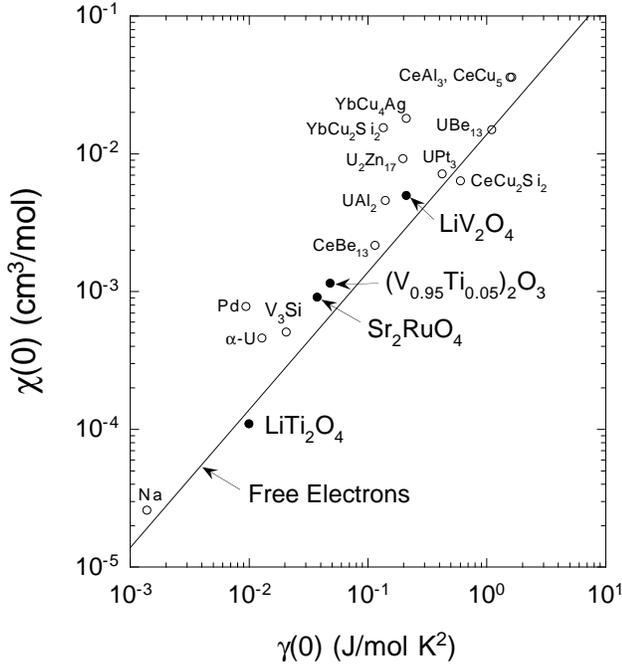}
\vglue 0.1in
\caption{Log-log plot of the magnetic susceptibility $\chi(0)$ versus electronic
specific heat coefficient $\gamma(0)$ at zero temperature for a variety of
$f$-electron heavy fermion and intermediate valence compounds compiled from the
literature (after Ref.~\protect\onlinecite{Lee1987}).  The plot also includes
data for several elemental and/or $d$-electron metals and our data point for
LiV$_2$O$_4$.  Here, a ``mol'' in the axis labels refers to a mole of transition
metal atoms for the $d$-metal compounds, and to a mole of $f$-electron atoms for
compounds containing lanthanide or actinide atoms.  The straight line
corresponds to a Wilson ratio $R_{\rm W} = 1$ for quasiparticles with spin $S =
1/2$ and $g$-factor
$g = 2$, which is also the Wilson ratio for a free-electron Fermi
gas.\label{FigChiVsGam}}
\end{figure}

For two high-magnetic-purity LiV$_2$O$_4$ samples~3 and~6, the electronic
specific heat coefficients $\gamma(T) \equiv C_{\rm e}(T)/T$ were found to be
$\gamma(1\,$K$) = 0.42$ and 0.43 J/mole\,K$^2$, respectively.  To our
knowledge, these values are significantly larger than previously reported for any
metallic transition metal compound.\cite{Ballou1996}  For LiTi$_2$O$_4$, we found
$\gamma = 0.018$\,J/mole\,K$^2$.  $\gamma(T)$ of LiV$_2$O$_4$ decreases rapidly
with increasing temperature from 4 to 30\,K and then decreases much more slowly
from a value of 0.13\,J/mol\,K$^2$ at 30\,K to 0.08\,J/mol\,K$^2$ at 108\,K\@. 
Even these latter two
$\gamma$ values are exceptionally large for a metallic $d$-electron compound. 
The temperature dependences of $\gamma$, $\chi$, the low-$T$ resistivity and the
$^7$Li NMR properties are remarkably similar to those of the heaviest mass
$f$-electron heavy fermion compounds.\cite{HFIV}  In a plot of $\chi(0)$ versus
$\gamma(0)$, the data point for LiV$_2$O$_4$ sits amid the cluster formed by the
$f$-electron heavy fermion and intermediate valent compounds as shown in
Fig.~\ref{FigChiVsGam},\cite{Lee1987} where several 
\noindent
data for elemental metals,
the A-15 superconductor V$_3$Si ($T_{\rm c} = 17$\,K),\cite{Junod1971,Maita1972}
and superconducting and/or metallic $d$-metal oxides LiTi$_2$O$_4$ ($T_{\rm c}
\leq 13.7$\,K),\cite{Johnston1976} Sr$_2$RuO$_4$ ($T_{\rm c}$ =
1\,K),\cite{Maeno1997} and (V$_{0.95}$Ti$_{0.05})_2$O$_3$,\cite{McWhan1971} are
also included for comparison.

From our theoretical modeling in Sec.~\ref{SecTheory}, Fermi liquid models and
the
$S = 1/2$  Kondo model (with a Fermi liquid ground state) are capable of
describing our
$C_{\rm e}(T)$ data for LiV$_2$O$_4$ from 1\,K up to $\sim 10$\,K, although the
magnitudes of the derived parameters remain to be understood theoretically.  The
localized moment model in Sec.~\ref{SecHTS} failed both qualitatively and
quantitatively to describe the data.  None of the models we used can account for
the additional contribution to $C_{\rm e}(T)$ at higher temperatures, from $\sim
10$\,K up to our high temperature limit of 108\,K, which appears to be distinct
from the contribution beginning at much lower $T$ and could arise from
orbital,\cite{Takigawa1996,Bao1997} charge and/or
spin\cite{Silva1996,Andrei1995} excitations.  The crystalline electric field
and/or the spin-orbit interaction may produce some energy level structure which
is thermally accessible within our temperature range.\cite{Radwanski1998} 
Conventional band structure effects cannot give rise to our
results.\cite{Harmon1998}

As is well-known for conventional metals, the electron-phonon interaction
increases $\gamma$ by the factor $(1 + \lambda)$, where $\lambda$ is the
electron-phonon coupling constant, but does not affect $\chi$; {\it i.e.}, ${\cal
D}^*(E_{\rm F}) \rightarrow {\cal D}^*(E_{\rm F}) (1 + \lambda)$ in
Eq.~(\ref{EqDOS:a}).  One can correct  the observed Wilson ratio for
electron-phonon interactions by multiplying the observed value by
$(1 +\lambda)$.\cite{Fulde1988}  The electron-phonon interaction is not taken
into account in any of the analyses or modeling we have done.  This correction
would have had a significant quantitative impact on our analyses if we used,
{\it e.g.}\ $\lambda \approx 0.7$ as in LiTi$_2$O$_4$ (Refs.\
\onlinecite{Heintz1989,McCallum1976}); most previous analyses of the specific
heats of other ($f$-electron) HF compounds also did not take the electron-phonon
interaction into account.\cite{HFIV}

From our combined specific heat and thermal expansion measurements on the same
sample~6 of LiV$_2$O$_4$ from 4.4 to 108\,K, we derived the Gr\"uneisen parameter
$\Gamma(T)$ which shows a dramatic enhancement below $\sim 25$\,K as the compound
crosses over from the quasilocal moment behavior at high temperatures to the
low-temperature Fermi liquid regime, confirming the discovery of Chmaissem {\it
et al.}\ from neutron diffraction measurements.\cite{Chmaissem1997}  Our
estimated extrapolated value of the electronic Gr\"uneisen parameter
$\Gamma_{\rm e}(0)$ is about 11.4, which is intermediate between values for
conventional metals and for $f$-electron heavy fermion compounds.  This large
value indicates a much stronger dependence of the mass-enhanced density of
states on the volume of the system than simply due to the decrease in the Fermi
energy with increasing volume as in the quasi-free electron picture.  In the
$f$-electron HF systems, the large $\Gamma_{\rm e}(0)$ values are thought to
arise from a strongly volume dependent hybridization of the $f$-electron
orbitals with those of the conduction
electrons.\cite{deVisser1989,Edelstein1983}  In the present case of
LiV$_2$O$_4$, the origin of the large $\Gamma_{\rm e}(0)$ is unclear.

It is conceivable that the same mechanism is responsible for the heavy fermion
behavior in LiV$_2$O$_4$ as in the $f$-electron heavy fermion systems if one of
the 1.5 $d$-electrons/V atom is localized on each V  atom due to electron
correlation effects and crystalline electric field orbital energy level
structure,\cite{Goodenough1997} and if the orbital occupied by the localized
electron is hybridized only weakly with the conduction electron states.  That
such localization can occur in similar systems is supported by calculations for
the $d^1$ compound NaTiO$_2$.\cite{Ezhov1998}  Additional scenarios for the
heavy fermion behavior mechanism(s) are given by Kondo {\it et al.}\cite{Kondo1997,Kondo1998} involving the geometric frustration for AF
ordering within the V sublattice and/or low-lying coupled dynamical
orbital-charge-spin excitations.  Further experimental and theoretical
investigations of the physical properties of LiV$_2$O$_4$ may thus reveal
interesting new physics which may also allow a deeper understanding of the
$f$-electron heavy fermion class of materials.

\section*{Acknowledgments} We are indebted to F.~Izumi for helpful communications
regarding the Rietveld analyses and to A.~Jerez and N.~Andrei for providing
high-accuracy numerical values\cite{Jerez1997} of the magnetic susceptibility
and specific heat of the $S = 1/2$ Kondo model.  We thank V.~Antropov, F.~Borsa,
O.~Chmaissem, J.~B.~Goodenough, R.~J.~Gooding, B.~N.~Harmon, J.~D.~Jorgensen,
M.~B.~Maple and A.~J.~Millis for helpful discussions and correspondence, and
V.~Antropov and B.~N.~Harmon for communicating to us the results of their
unpublished band structure calculations for LiV$_2$O$_4$.  Ames Laboratory is
operated for the U.S. Department of Energy by Iowa State University under
Contract No.\ W-7405-Eng-82.  This work was supported by the Director for Energy
Research, Office of Basic Energy Sciences.

\end{document}